\documentclass[%
 reprint,
 superscriptaddress,
 amsmath,amssymb,
 aps,
 prx
]{revtex4-2}

\usepackage{braket}
\usepackage{graphicx}
\usepackage{dcolumn}
\usepackage{bm}
\usepackage{siunitx}
\usepackage{xcolor} 
\usepackage{tabularx}
\usepackage{dsfont}					
\usepackage{mathrsfs}				
\usepackage{float}
\usepackage{hyperref}

\newcommand{\comment}[1]{}

\newcommand{\be}[0]{\begin{equation}}	
\newcommand{\ee}[0]{\end{equation}}

\renewcommand{\Im}{\mathrm{Im}}

\usepackage{hyperref}
\hypersetup{colorlinks, 
	linkcolor={blue!75!black!80!yellow},
	citecolor={blue!75!black!80!yellow}, 
	urlcolor={blue!75!black!80!yellow}
	}

\usepackage[normalem]{ulem}

\begin{document}



\title{Probing the Berezinskii-Kosterlitz-Thouless vortex unbinding transition\\
in two-dimensional superconductors using local noise magnetometry}


\author{Jonathan~B.~Curtis}
\email[]{jon.curtis.94@gmail.com}
\affiliation{College of Letters and Science, University of California, Los Angeles, CA 90095, USA}
\affiliation{Department of Physics, Harvard University, Cambridge, MA 02138, USA}
\affiliation{Institute for Theoretical Physics, ETH Z{\"u}rich, Z{\"u}rich, CH}
\author{Nikola~Maksimovic}
\author{Nicholas~R.~Poniatowski}
\author{Amir~Yacoby}
\author{Bertrand~Halperin}
\affiliation{Department of Physics, Harvard University, Cambridge, MA 02138, USA}

\author{Prineha~Narang}
\affiliation{College of Letters and Science, University of California, Los Angeles, CA 90095, USA}
\author{Eugene~Demler}
\affiliation{Institute for Theoretical Physics, ETH Z{\"u}rich, Z{\"u}rich, CH}

\date{\today}

\begin{abstract}
The melting of quasi-long-range superconductivity in two spatial dimensions occurs through the proliferation and unbinding of vortex-antivortex pairs -- a phenomenon known as the Berezinskii-Kosterlitz-Thouless (BKT) transition.
Although signatures of this transition have been observed in bulk measurements, these experiments are often complicated, ambiguous, and unable to resolve the rich physics of the vortex unbinding transition. Here we show that local noise magnetometry is a sensitive, noninvasive probe that can provide direct information about the scale-dependent vortex dynamics.
In particular, by resolving the distance and temperature dependence of the magnetic noise, it may be possible to experimentally study the renormalization group flow equations of the vortex gas and track the onset of vortex unbinding {\it in situ}.  Specifically, we predict i) a nonmonotonic dependence of the noise on temperature and ii) the local noise is almost independent of the sample-probe distance at the BKT transition.
We also show that noise magnetometry can distinguish Gaussian superconducting order-parameter fluctuations from topological vortex fluctuations and  can detect the emergence of unbound vortices.
The weak distance dependence at the BKT transition can also be used to distinguish it from quasiparticle background noise. 
Our predictions may be within experimental reach for a number of unconventional superconductors.
\end{abstract}

\maketitle

\section{Introduction}
The study of critical phenomena in low-dimensional systems is rich and complex, in part due to the increased importance of fluctuations in these systems.
This is perhaps most clearly manifested in the Mermin-Wagner-Hohenberg theorem~\cite{Mermin.1966,Hohenberg.1967,Coleman.1973}, which shows that spontaneous breaking of a continuous symmetry group is impossible in dimensions two and lower due to long-wavelength fluctuations of the order parameter.
In particular, long-range superfluid order, which is characterized by the spontaneous breaking of a $U(1)$ symmetry, is therefore impossible in two-dimensions as a matter of principle.
However, in a series of ground-breaking works by Berezinskii~\cite{Berezinskii.1971} and Kosterlitz and Thouless~\cite{Kosterlitz.2002,Kosterlitz.2016}, it was shown that the situation is more nuanced.

In particular, while true long-range superfluid order is indeed impossible in two dimensions, a ``quasi-long-range ordered" (QLRO) phase is possible, and this is sufficient to enable superfluid transport and macroscopic quantum coherence effects~\cite{Mikeska.1970}.
The key distinction between the QLRO phase and disordered phase is whether vortices\textemdash topological defects in the order parameter which possess quantized angular momentum\textemdash are bound or free.
At low temperatures, vortices and antivortices experience an attractive force and form tightly-bound inert pairs.
In contrast, at high-temperatures these pairs disassociate due to thermal agitation and become unbound and free to wander, leading to dissipation and phase slips which ruin superfluidity.
The transition between these two phases, known as the Berezinskii-Kosterlitz-Thouless (BKT) transition, is therefore characterized by the nature of the vortex-antivortex interaction and has no local order parameter, circumventing the Mermi-Wagner-Hohenberg theorem~\cite{Mermin.1966,Hohenberg.1967,Coleman.1973} and leading to its classification as a ``topological phase transition."

There are a number of striking predictions for the nature of this transition, including a universal jump in the renormalized superfluid density $\rho^*_{\rm 2D}(T)$ which jumps from zero for $T > T_{\rm BKT}$ to $\rho_{\rm 2D}^*(T_{\rm BKT}^-) = \frac{2}{\pi}T_{\rm BKT}$ upon crossing the transition~\cite{Nelson.1977}.
Another notable prediction is the divergence of the correlation length $\xi_+(T)$ upon approaching the transition from above, which exhibits an essential singularity as 
$\xi_+(T)\sim\exp[ b (T/T_{\rm BKT}- 1)^{-1/2}]$ (with $b$ a constant) diverging faster than any power law~\cite{Kosterlitz.2001}.
A number of these predictions have been verified in a wide range of realizations, including ultracold quantum gases~\cite{Hadzibabic.2006,Hadzibabic.2013}, establishing the validity of the topological BKT transition~\cite{Kosterlitz.2016} across a variety of physical realizations.

Though originally formulated in the context of neutral superfluids~\cite{Kosterlitz.2002}, it was quickly shown that the BKT transition is also relevant for two-dimensional superconductors~\cite{Beasley.1979,Halperin.1979,Doniach.1979}, provided the sample dimensions are sufficiently small compared to the Pearl length~\cite{Pearl.1964}.
In this case the BKT transition directly manifests in the bulk electrical transport properties~\cite{Halperin.1979}, which has enabled confirmation of the BKT theory in a number of samples, including recently to a very high degree of accuracy in NbN~\cite{Weitzel.2023}.
Owing to the charged nature of the condensate, vortices not only carry angular momentum but also magnetic flux, and therefore directly affect the magnetodynamic response of superconductors. As such, the BKT transition can also manifest in global flux noise experiments~\cite{Kim.1999,Houlrik.1994,Korshunov.2002,Shaw.1996,Festin.1999,Bjornander.1996,Rogers.1992}, which detect the thermal motion of magnetic vortices and allow one to make inferences about vortex interactions.

To date however, existing experimental probes of BKT physics in materials have been limited to probing bulk long-wavelength properties. 
As such, these approaches can only indirectly infer physics on the length-scale of individual vortices, which is at the heart of the BKT theory. In addition, it can be challenging to unambiguously identify the BKT transition from global properties alone, as it can easily be obscured by long-wavelength disorder, material inhomogeneity~\cite{Mondal.2011,Benfatto.2008,Benfatto.2009}, and heating effects~\cite{Weitzel.2023}.
Background contributions to the conductivity from both fermionic quasiparticles~\cite{Minnhagen.1987} and Gaussian superconducting fluctuations~\cite{Larkin.2005} can also further obscure BKT physics.

Meanwhile, it has become increasingly more important to understand how BKT physics manifests in two-dimensional superconductors due to recent advances in the discovery and design of atomically thin Van der Waals superconductors.
This includes a large number of unconventional superconductors such as BSCCO~\cite{Yu.2019}, FeSe~\cite{Faeth.2021}, WTe$_2$~\cite{Fatemi.2018,Sajadi.2018}, as well as a number of graphene allotropes~\cite{park2021tunable,Cao.2018,Zhou.2021,Zhang.2023}, including putative fluctuating triplet superconductivity~\cite{Curtis.2023}.
Understanding the details of the BKT transition in these atomically thin and highly tunable structures might shed some light onto the microscopic processes that drive superconductivity in these materials.

\begin{figure}
    \centering
    \includegraphics[width=\linewidth]{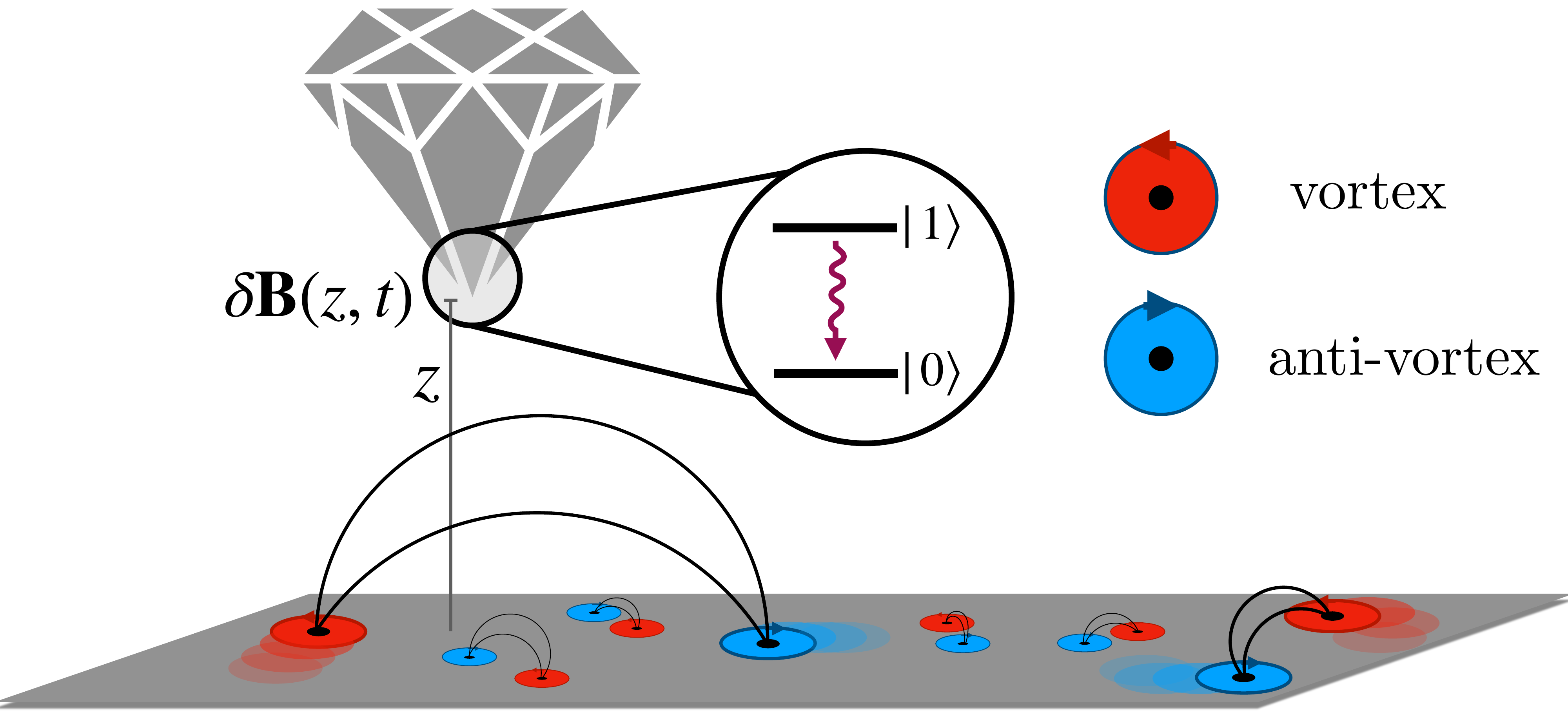}
    \caption{Schematic overview of the proposal. 
    A two-dimensional superconductor hosts vortices which exhibits a BKT phase transition.
    This results in a characteristic scale-dependent magnetic noise, which can be studied via local magnetometry, both via $1/T_1$- and $1/T_2$-like measurements.}
    \label{fig:schematic}
\end{figure}

In this work, we show that local noise magnetometry can in principle give access to the full scale-dependent vortex dynamics across the BKT transition in a non-invasive way -- see Fig.~\ref{fig:schematic}.
Most notably, we find a maximum in the local magnetic noise at the BKT transition as a function of temperature.
This local noise maximum is nearly independent of sample-probe distance at sufficiently low frequencies.
At zero frequency this scale-invariance will persist up to distances which become comparable to either the sample size, Pearl length~\cite{Pearl.1964}, or other macroscopic cutoff scale~\footnote{At finite frequency there is an additional cutoff scale at $\ell_\omega \sim \sqrt{D/\omega}$ above which the frequency dependence of vortex diffusion becomes relevant.}.
We will show later that this scale-invariance is crucial, as it enables the disentangling of vortex and quasiparticle effects. 

Above, but close to, the transition temperature, we predict signatures of finite-size crossover effects, which can be studied {\it locally} and {\it in situ}; at even higher temperatures further from the transition, an additional contribution to the magnetic noise from unbound free vortices can be discerned and used to study the vortex proliferation. 
We also show that the magnetic noise dependence on the sample-probe distance can be used to distinguish Gaussian superconducting fluctuations from vortex effects, further demonstrating the power of local noise magnetometry for studying the BKT transition.

This work is motivated in part by recent progress in the development of atom-like solid-state defects, nitrogen-vacancy (NV) centers in diamond~\cite{Casola.2018,Rovny.2024} in particular, that can sense weak magnetic fields at the nanoscale.  
To date, NV noise magnetometry has been experimentally employed to study a variety of many-body systems, including magnons in ferromagnetic insulators~\cite{Du.2017}, magnon hydrodynamics in a Van der Waals material~\cite{Xue.2024}, Johnson-Nyquist noise of electrons in a metal~\cite{Kolkowitz.2015}, the electron-phonon Cherenkov effect in a driven graphene~\cite{Andersen.2019}, and fluctuations in antiferromagnetic topological insulator MnBi$_2$Te$_4$~\cite{McLaughlin.2022}. On the theory side, NVs have been argued to be a useful probe for a far wider array of phenomena including the study of quasiparticles~\cite{Dolgirev.2022,Chatterjee.2022} and stripes~\cite{Konig.2020} in  atomically-thin superconductors, two-dimensional electronic Wigner crystals~\cite{Dolgirev.2023}, one- and two-dimensional quantum phases~\cite{Rodriguez-Nieva.2018,Agarwal.2017}, critical magnetic fluctuations~\cite{Machado.2022}, semimetals~\cite{Zhang.2022o8f}, and correlated magnetic insulators~\cite{Chatterjee.2019,Khoo.2022}.
Additionally, it has even been proposed that multiple NV centers could be used to cooperatively probe materials via entangled qubit states~\cite{Li.20238nq}. 
For a comprehensive overview of the capabilities of NV sensing see, e.g. Ref.~\onlinecite{Rovny.2024}.

The remainder of this paper is structured as follows.
In Sec.~\ref{sec:probing} we show how the local magnetic noise can be used to directly map out the frequency and momentum dependence of the vortex-antivortex interactions.
Then in Sec.~\ref{sec:results} we present a detailed analysis of this dependence and identify a number of key predictions. 
In Sec.~\ref{sec:AL} we compare these predictions against those based on a Gaussian model of Aslamazov-Larkin type fluctuations.
In Sec.~\ref{sec:expt} we survey a number of relevant materials which may be promising for detecting BKT physics and discuss possible complications due to quasiparticles. 
Finally, in Sec.~\ref{sec:conclusion} we conclude with a discussion of future directions and challenges. 



\section{Vortex Interactions and magnetic noise}
\label{sec:probing}

In this section, we relate the magnetic field noise at a distance $z$ away from the superconducting sample (see Fig.~\ref{fig:schematic}) to the vortex correlation functions. To obtain the noise spectrum we assume the superconductivity is truly two-dimensional, i.e., i) the sample thickness is much smaller than the penetration depth and ii) the BKT transition temperature $T_{\rm BKT}$ is much less than the pairing temperature $T_{\rm BCS}$ -- corresponding to the London limit and allowing us to neglect effects due to quasiparticles~\cite{Chatterjee.2022,Dolgirev.2022} (for a more precise discussion see Sec.~\ref{sub:qp}). 
Assumption (i) is already realized in a number of conventional and unconventional superconductors and is especially relevant for Van der Waals materials such as FeSe, graphene, and NbSe$_2$ as well as thin-films of NbN and high-$T_c$ cuprates.
While it is unclear whether assumption (ii) can be truly realized, predictions of our theory should still be relevant in any situation where there is a significant separation between $T_{\rm BKT}$ and the three-dimensional transition temperature, as discussed in Sec.~\ref{sec:expt}.

The magnetic noise spectrum at a distance $z$ above the sample and at a frequency $\omega$ which satisfies the magnetostatic condition $\omega \ll c/z$ can be directly related to the equilibrium fluctuations of the transverse current -- see Appendix~\ref{app:magnetostatic}:
\begin{multline}
\label{eqn:noise}
\mathcal{N}_{zz}(\omega,z) = \int dt \, e^{i\omega t} \langle B_z(z,t) B_z(z,0) \rangle \\
=  (\mu_0/2)^2 \int_{\bf q} e^{-2zq} S^\perp(\omega,q),
\end{multline}
with $\mu_0$ the vacuum permeability.
For vortex fluctuations in the superconductor, the correlation function $S^\perp(\mathbf{q},\omega)$ can be related to the vortex structure factor 
\begin{equation}
    \label{eqn:chi-q}
    \chi(\omega,q) = \int dt\, e^{i\omega t}\langle n(t,\mathbf{q})n(0,-\mathbf{q}) \rangle
\end{equation} 
via
\begin{equation}
    \label{eqn:S-BKT}
    S^\perp(\omega,q) = (2e)^2 (2\pi \rho_{\rm 2D})^2\frac{\chi(\omega,q)}{q^2} .
\end{equation}
Here $\rho_{\rm 2D}$ is the bare two-dimensional superfluid density at low temperatures, and 
\begin{equation}
    n(t,\mathbf{q}) =  \sum_j n_j \exp\left(i \mathbf{q}\cdot\mathbf{R}_j(t) \right) 
\end{equation}
is the two-dimensional vortex density, with $n_j = \pm 1$ and the overall ``neutrality" constraint $\sum_j n_j = 0$ (in analogy with the Coulomb plasma model).
Therefore, by characterizing the local magnetic noise spectrum we can directly read out the vortex dynamic structure factor as we tune through the BKT phase transition. 

It is convenient to express the noise in terms of the vortex dielectric function, $\epsilon_v(\omega,{q})$, which characterizes the dynamically screened vortex-vortex interactions.
We emphasize that $\epsilon_v(\omega,{q})$ is the dielectric function for the effective vortex-vortex Coulomb interaction and is not to be confused with the physical dielectric function, which is not relevant for the transverse current fluctuations and will not be discussed in this paper.
As per the standard relations between charge susceptibility and dielectric function, and employing the fluctuation-dissipation relation~\cite{Kim.1999,Houlrik.1994}, we can write 
\begin{equation}
     \chi(\omega,q)  = - \frac{2T}{\omega}\frac{1}{v({q})} \Im \Big[ \frac{1}{\epsilon_v(\omega,q)}\Big]
\end{equation}
where $v({q})=  4\pi^2 \rho_{\rm 2D}/{q}^2$ is the bare interaction between vortex charges.
The current noise spectrum is then given by: 
\begin{equation}
\label{eqn:S-epsilon}
S^\perp(\omega,q) = 2 (2e)^2 T \rho_{\rm 2D} \Im \Big[\frac{-1}{\omega \epsilon_v(\omega,q)} \Big],
\end{equation}
and the corresponding magnetic noise spectrum reads
\begin{align}
\mathcal{N}_{zz}(\omega,z) = 2 N & (\mu_0  e/\pi )^2   T T_{\rm BKT}^{(0)} \notag\\
& \times \int_0^{\infty} dq \, q \,e^{-2zq} \, \Im \Big[\frac{-1}{\omega \epsilon_v(\omega,q)} \Big] .\label{eqn:noise-epsilon}
\end{align}
Here we have introduced $T_{\rm BKT}^{(0)} = \frac{\pi}{2}\rho_{\rm 2D}$, which is the BKT transition temperature obtained by assuming that the universal superfluid density jump~\cite{Nelson.1977} is applicable to the bare superfluid density $\rho_{\rm 2D}$, obtained say at zero temperature. 
We have also included the prefactor $N$, which corresponds to considering a stack of $N$ equally spaced, uncoupled layers separated by a distance $a\ll z$ -- having several such layers makes the noise larger and, thus, more feasible experimentally (see Sec.~\ref{sec:expt}). Below, we will present results per single layer allowing us, thus, to drop the parameter $N$ from our analysis (i.e. $N = 1$).

We will further focus on the experimentally most relevant case of classical diffusive motion of vortices following the Bardeen-Stephens model~\cite{Bardeen.1965,Halperin.1979}, with the vortex mobility $\mu$ and diffusion constant $D = \mu T$.
More elaborate forms of vortex motion as well as quantum effects are left to future work. We evaluate the vortex dielectric function $\epsilon_v(\omega,q)$ for this model in the following section.

For future reference, we turn to discuss relevant dimensionful parameters that determine the magnetic noise properties as well as experimental feasibility of our proposal, which we further examine in Sec.~\ref{sec:expt}. We first note that
the typical frequency associated with the vortex dynamics is expected to scale as $\omega \sim D q^2$ so that the integration measure $qdq/\omega \sim 1/D$. The reference diffusion constant $D(T)$ is referenced at the BKT transition temperature (the vortex mobility is roughly temperature independent):
\begin{equation}
    D(T_{\rm BKT}) \equiv D_0 = \mu T_{\rm BKT}.
\end{equation}
The noise scale is then referenced with respect to the overall scale 
\begin{equation}
    \mathcal{N}_0 = 2(\mu_0 e/\pi)^2 \frac{(T_{\rm BKT})^2}{D_0 } = 2(\mu_0 e/\pi)^2 \frac{T_{\rm BKT}}{\mu}.
    \label{eq:noise_scale}
\end{equation}
Normalizing by $\mathcal{N}_0$, we rewrite  Eq.~\eqref{eqn:noise-epsilon} as
\begin{equation}
\frac{\mathcal{N}_{zz}(\omega,z)}{\mathcal{N}_0} = D \frac{T_{\rm BKT}^{(0)}}{T_{\rm BKT}}\int_0^{\infty} dq\, q \,e^{-2zq} \, \Im \Big[\frac{-1}{\omega \epsilon_v(\omega,q)} \Big].
\end{equation}
Within the theoretical model used here the ratio $T_{\rm BKT}^{(0)}/T_{\rm BKT} = \rho_{\rm 2D}/\rho_{\rm 2D}^* = \epsilon_c$ where $\epsilon_c$ is the value of the vortex dielectric constant at zero frequency and momentum just below the transition temperature; that is, $\epsilon_c = \lim_{\omega,q\to 0} \lim_{T \to T_{\rm BKT}^{-}} \epsilon_v(\omega,q)$.
We can therefore rewrite this as 
\begin{equation}
\label{eqn:noise-epsilon_v2}
\frac{\mathcal{N}_{zz}(\omega,z)}{\mathcal{N}_0} = D \int_0^{\infty} dq\, q \,e^{-2zq} \, \Im \Big[\frac{-\epsilon_c}{\omega \epsilon_v(\omega,q)} \Big].
\end{equation}
We have also ignored any reduction of the superfluid density due to quasiparticle excitations. 
Length scales are referenced with respect to the coherence length $\xi_c$ which physically corresponds to the size of the vortex core and serves as an ultraviolet cutoff on the validity of the London theory for vortex motion. 
In our treatment this is taken as an input to the model which will be approximated by the value of the coherence length in the $ab$ plane at low temperatures.
Using $\xi_c$ we can introduce the reference frequency 
\begin{equation}
    \omega_0 = D_0 \xi_c^{-2} = \mu T_{\rm BKT}\xi_c^{-2}.
\end{equation}
Since the vortex coherence length $\xi_c$ is the smallest length scale in the problem, $\omega_0$ will be the largest frequency scale in the problem.

\section{Results and Discussion}
\label{sec:results}
The key question of interest is the behavior of the effective dielectric function $\epsilon_v(\omega,q)$.
The behavior of this in certain limits is well known, and in general should behave similar to a real (charge) dielectric with a corresponding mapping from vortex parameters to charge parameters. 
That is, at low temperatures, $\epsilon_v(\omega,{q})$ should resemble the dielectric function of a ``vortex dielectric," while at high temperatures $\epsilon_v$ will behave as that of a ``vortex plasma." 

This behavior is quantitatively described by the solution of the Kosterlitz renormalization group (RG) equations~\cite{Kosterlitz.2001,Young.2000,Jose.1977}, which yield the effective vortex interaction $\epsilon_v^{-1}$ and fugacity $y$ at a particular length scale $\ell$. 
In the limit of small fugacity $y\ll 1$, these read 
\begin{subequations}
\label{eqn:rg-eqn}
    \begin{align}
        & \frac{d\epsilon_v(\ell)}{d\log \ell} = 4 \pi^3 y^2(\ell)\\
        & \frac{d y(\ell)}{d\log \ell} = 2\left(1 - \frac{T_{\rm BKT}^{(0)}}{T} \frac{1}{\epsilon_v(\ell)} \right) y(\ell).
    \end{align}
\end{subequations}
In addition, we specify the microscopic value of these parameters, with $\epsilon_v(\xi_c) = 1$ and $y(\xi_c)=y_0  = c \exp(-T_{\rm BKT}^{(0)}/T)$ with $c = 0.1$ used here.
Note the exact behavior of the results here are somewhat dependent on the model for used for the bare vortex fugacity.
Due to screening of vortex interactions the actual transition temperature is slightly shifted from the bare value $T_{\rm BKT}^{(0)}$ to $T_{\rm BKT}\approx 0.8117T_{\rm BKT}^{(0)}$ (see Appendix~\ref{app:BKT-temp}).
This corresponds to a value of $\epsilon_c = 1.2320$.

\begin{figure}
    \centering
    \includegraphics[width=\linewidth]{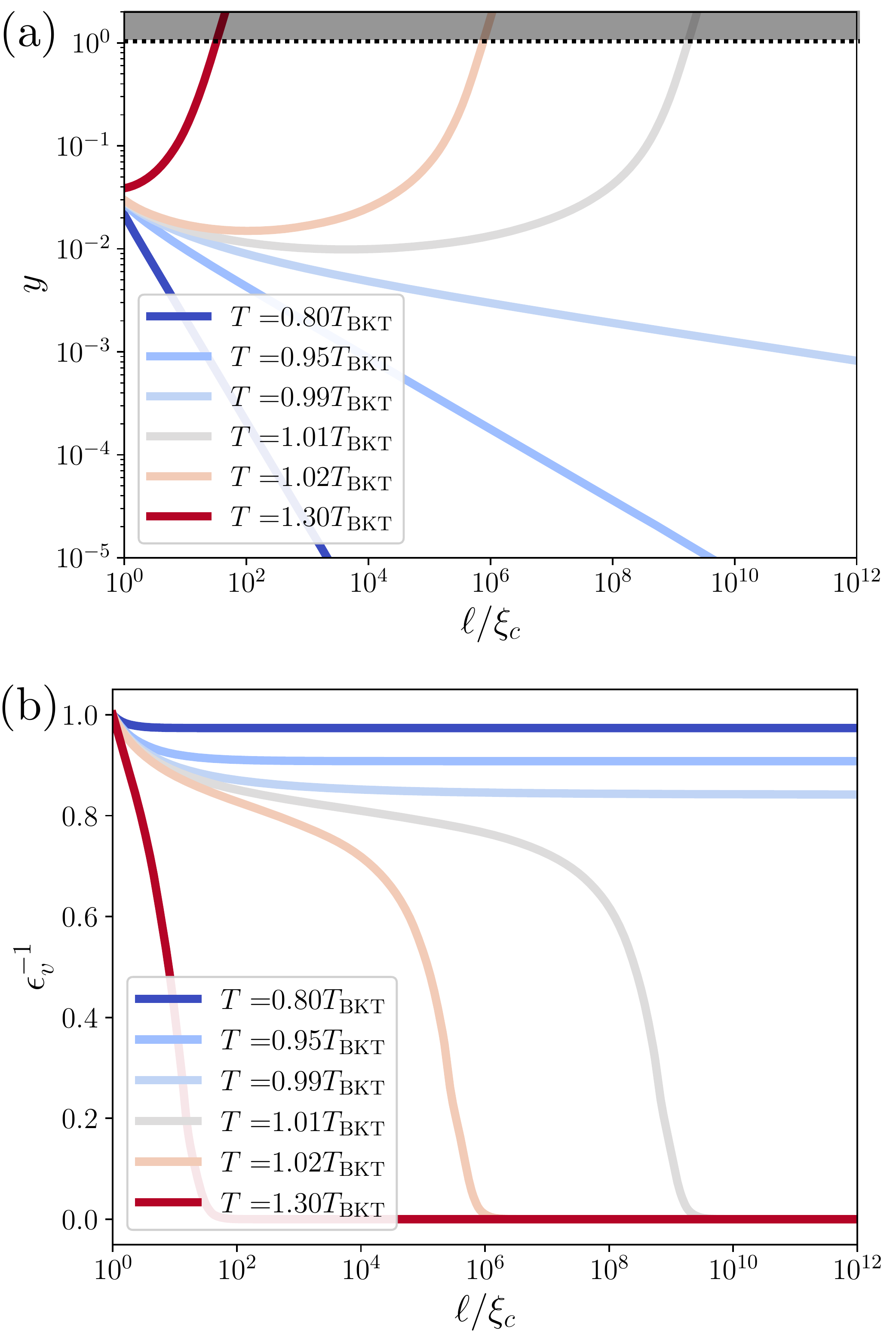}
    \caption{(a) Vortex fugacity $y$ flow as a function of length scale $\ell$ for different temperatures. 
    At low temperatures the fugacity flows to zero, while at high temperatures it diverges to the non-perturbative regime of $y > 1$ (shaded area). 
    (b) Static inverse vortex dielectric constant $\epsilon_v^{-1}$ flow as as function of $\ell$.
    Below the transition $\epsilon_v^{-1}$ remains finite as $\ell \to \infty$, while at high temperatures this exhibits a crossover before converging to $\epsilon_v^{-1} = 0$. 
    } 
    \label{fig:rg}
\end{figure}

In Fig.~\ref{fig:rg} we present the solutions to the RG equations for (a) the vortex fugacity and (b) the static vortex dielectric constant as a function of length scale $\ell$ for different values of the temperature $T$.
At low temperatures ($T < T_{\rm BKT}$) the vortex fugacity is irrelevant and thus upon increasing $\ell$ we see $y\to 0$ while, simultaneously the vortex dielectric constant $\epsilon_v^{-1}$ tends to a finite value reflecting the finite renormalized phase stiffness $\rho_{\rm 2D}^* = \rho_{\rm 2D}\epsilon_v^{-1}(\ell=\infty) > 0$.

\begin{figure}
    \centering
    \includegraphics[width=\linewidth]{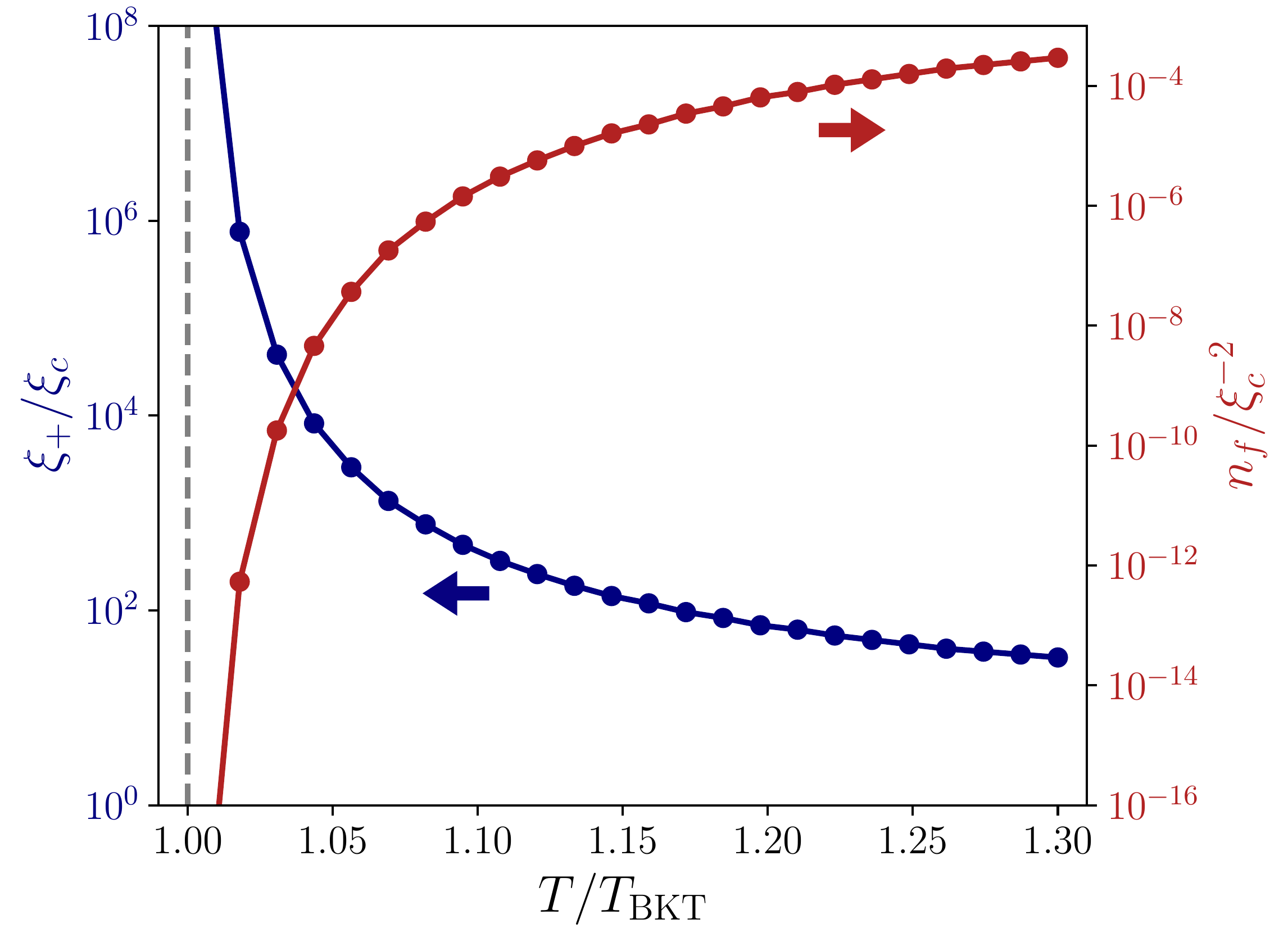}
    \caption{Left axis: length scale at which vortices become unbound, $\xi_+(T)/\xi_c$.
    This diverges such that $\log\xi_+ \sim 1/\sqrt{T - T_{\rm BKT}}$, which grows very rapidly upon approaching the transition from above. 
    Right axis: density of free vortices above $T_{\rm BKT}$ obtained as $n_f = \xi_+^{-2}/\pi$. } 
    \label{fig:xi-nf}
\end{figure}

At high temperatures ($T > T_{\rm BKT}$) the vortex fugacity instead becomes a relevant perturbation and thus increases with increasing $\ell$ until it reaches the nonperturbative regime of $y \gtrsim 1$.
At this point, the RG equations fail, and we interpret the scale $\ell =\xi_+$ at which this happens as the typical intervortex separation in the free vortex gas, with areal density $n_f = 1/(\pi \xi_+^2)$.
As it is a plasma, the vortex-vortex interactions become screened, and the vortex dielectric constant $\epsilon_v^{-1} \to 0$ captures this.
Since this is related to the renormalized phase stiffness by $\rho_{\rm 2D}^* = \rho_{\rm 2D} \epsilon_v^{-1} = 0$, this is seen to be a normal nonsuperconducting phase.
The length scale $\xi_+(T)$ is shown in Fig.~\ref{fig:xi-nf}, along with the corresponding free vortex density $n_f$, referenced in terms of the coherence length $\xi_c$. 
We can see that both exhibit essential singularities at $T = T_{\rm BKT}$, as expected. 

From knowledge of the scale-dependent vortex dielectric constant $\epsilon_v(\ell)$, it is possible to compute the dynamical dielectric function by essentially integrating the dynamical response from pairs over all length scales.
Previous studies have computed the long-wavelength dielectric function, $\epsilon_v(\omega,{q} = 0)$ using this procedure~\cite{Ambegaokar.1978,Ambegaokar.1979,Ambegaokar.1980}.
However, using NV relaxometry we are now in a position to probe the full spatial profile of the dielectric response function $\epsilon_v(\omega,{q})$.
In Appendix~\ref{app:finite-q-eps} we generalize previous long-wavelength results to obtain the finite-momentum and finite frequency vortex dielectric function $\epsilon_v(\omega,{q})$. 
We can write the result succinctly as 
\begin{multline}
\label{eqn:epsilon_wz}
    \epsilon_v(\omega,{q}) = 1 + \underbrace{\int_{\xi_c}^{\xi_+} d\ell \frac{d\epsilon_v(\ell)}{d\ell} \frac{F({q} \ell/2)}{1- i\frac{\omega \ell^2}{14D}}}_{\equiv \epsilon_b(\omega,q)\ (\textrm{bound vortices})}\\
    + \underbrace{\frac{4 \pi^2 \rho_{\rm 2D}n_f \mu}{D {q}^2 - i\omega}}_{\equiv \epsilon_f(\omega,q)\ (\textrm{free vortices})},
\end{multline}
where the filter function $F(x) = 2J_1(x)/x$ is approximately one for $\ell \ll q^{-1}$ and decays for $\ell \gg q^{-1}$, isolating the dominant contribution as arising from pairs with separation less than the wavelength being probed by $q$.
For numerical evaluation, this filter function will be replaced by a Gaussian $\tilde{F}(x) = \exp(-x^2/8)$ to remove the negative oscillations of $F(x)$ which can cause issues in the numerical integrals. 
See Appendix~\ref{app:finite-q-eps} for more information.
One can interpret the dynamical factor $\left(1- i\frac{\omega \ell^2}{14D}\right)^{-1}$ as encoding the diffusive relaxation rate of pairs at length scale $\ell$ to perturbations at frequency $\omega$, while the factor $\frac{d\epsilon_v}{d\ell}$ encodes the polarizeability of the pairs at that length scale. 
The factor of $14$ is known arise from the more careful solution of the dynamics of a bound votex-pair in a logarithmic potential, as found in Ref.~\onlinecite{Ambegaokar.1979}.

\begin{figure}
    \centering
    \includegraphics[width=\linewidth]{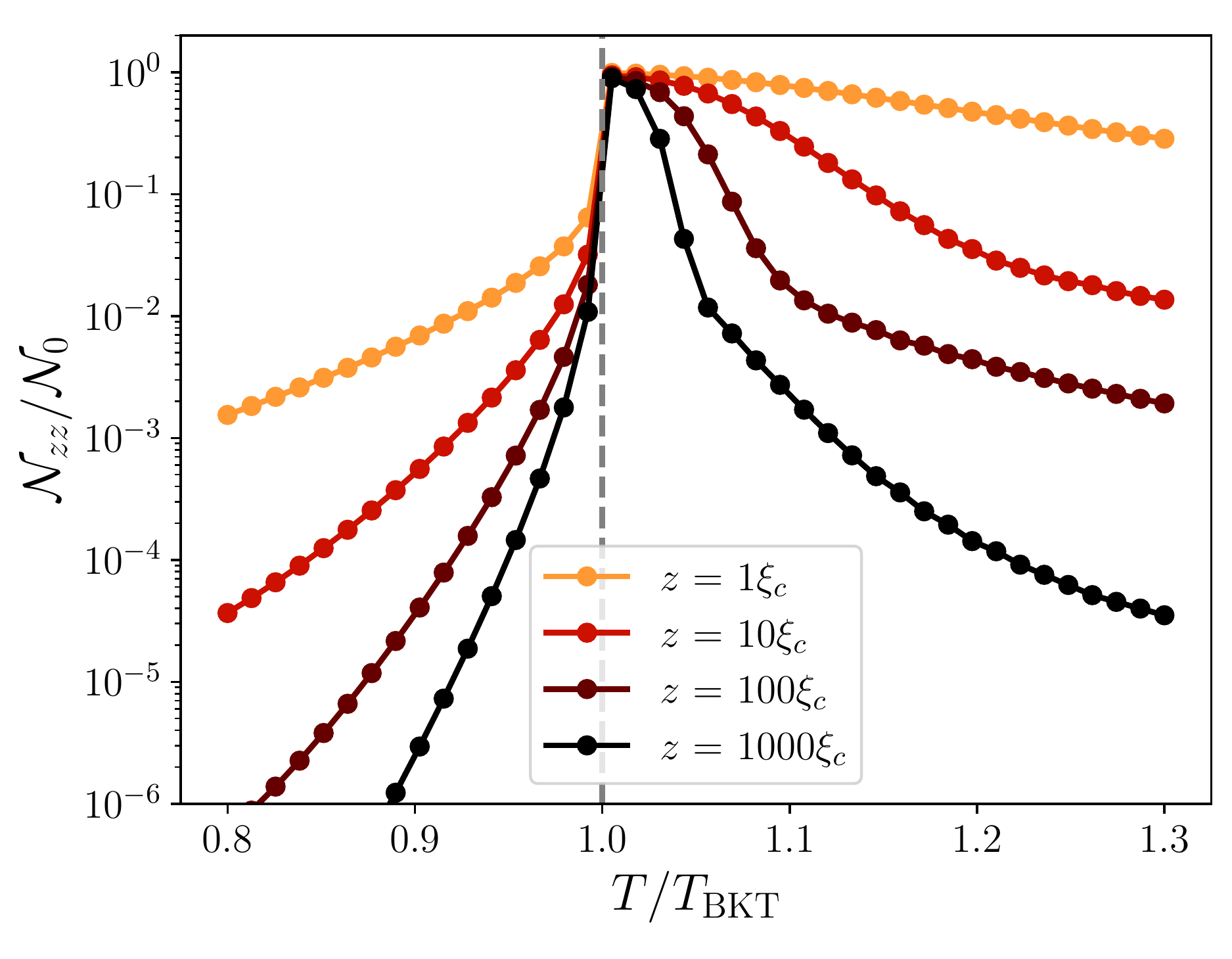}
    \caption{Local magnetic noise as a function of temperature for various qubit-sample distances $z$ near the BKT transition in the limit of $\omega \to 0$.
    The true transition is shown by the dashed line.
    }\label{fig:noise-T}
\end{figure}

For $T < T_{\rm BKT}$ the cutoff length scale $\xi_+ \to \infty$ and the Drude weight of free vortices vanishes as $n_f = 0$, so the contribution to the vortex dielectric function is purely due to bound pairs. 
However, for $T > T_{\rm BKT}$ the cutoff $\xi_+$ becomes finite, as does the finite density of free pairs with density $n_f \sim \xi_+^{-2}$.
These pairs then contribute as a Drude-type response which, at small momenta and low frequencies will become singular and dominant over the bound contribution. 
In general, even above the transition, there will also be a contribution from the bound pairs at length scales less than $\xi_+$, which may be large and thus it is important to include contributions from both the free vortex contribution as well as the bound pairs. 
Using the result of Eq.~\eqref{eqn:epsilon_wz} we arrive at the computed NV qubit noise spectrum $\mathcal{N}_{zz}$ as a function of temperature for different frequency and length scales.
The central result is encapsulated in Fig.~\ref{fig:noise-T}, which presents the low frequency $\omega \to 0$ limit of the magnetic noise spectrum $\mathcal{N}_{zz}$ for different NV-sample distances $z$ as a function of temperature $T$. 
In Fig.~\ref{fig:noise-T} the true transition temperature $T_{\rm BKT}$ is indicated by the vertical dashed line.

\subsection{Temperature Dependence}
\label{sub:temp}

We now comment on the important features of Fig.~\ref{fig:noise-T}.
Below the transition we see that the magnetic noise varies essentially as a power law in distance $z$ with a temperature dependent exponent that vanishes at $T_{\rm BKT}$. 
Right at the transition the magnetic noise is essentially scale invariant, and independent of qubit distance $z$ (up to logarithmic corrections), as might be expected near a phase transition.
A more complete discussion of the magnetic noise dependence on distance can be found in Sec.~\ref{sub:dist}, but the key result\textemdash the scale invariance of the noise at $T_{\rm BKT}$ can be easily understood based on dimensional analysis. 
The conductivity due to vortex motion, obtained in Appendix~\ref{app:conductivity}, is $\sigma^\perp_v(\omega,q) = \frac{\rho_{\rm 2D}}{-i\omega \epsilon_v(\omega,q)}$.
Based on diffusive motion, we expect the characteristic vortex frequency scale to be related to momentum by $i\omega \sim Dq^2$ and thus the transverse conductivity should scale at low frequencies as $\sigma^\perp_v(q)  \sim \frac{1}{Dq^2 \epsilon_v(q)}$.
In general, the vortex dielectric function $\epsilon_v(q)$ exhibits power law dependencies, but at the critical point we expect it to be essentially constant, owing to the marginality of the vortices at this point. 
Therefore the magnetic noise scales as 
\begin{equation}
    \mathcal{N}_{zz} \sim \int_0^\infty dq q e^{-2zq} \sigma^\perp_v(q) \sim \int_{q_{\rm min}}^\infty \frac{dq}{q} e^{-2zq}.
\end{equation}
Therefore this will be approximately scale independent as $z$ can be absorbed in to $dq/q$ (see Appendix~\ref{app:magnetostatic} for the magnetic noise as a function of the conductivity).
However, it is seen that for small $q$ this integrand is logarithmically divergent and therefore this scale independence is ultimately cutoff by some infrared cutoff\textemdash here denoted as $q_{\rm min}$\textemdash which will be set by the minimum of the system size, Pearl length~\cite{Pearl.1964}, or diffusive length $\ell_{\omega}$.

Returning to Fig.~\ref{fig:noise-T} we see that above the transition temperature we see a more complicated dependence, and in particular we see that the magnetic noise is nonmonotonic with temperature at fixed distance $z$, with the noise maximum occurring at the BKT transition. 
As the temperature continues to increase above $T_{\rm BKT}$ the magnetic noise then decreases\textemdash sharply at first\textemdash before reaching a pseudo-plateau occurring at a $z$-dependent temperature. 

The emergence of the pseudo-plateau is relatively easy to understand; it signals the qubit distance $z$ is large enough to probe the magnetic noise due to the free vortices, which will onset at lower temperatures for larger $z$.
The residual temperature dependence in this regime is mostly due to the temperature dependence of the free vortex density $n_f(T)$.

To understand the intermediate regime, it is important to note that, while for any temperature $T > T_{\rm BKT}$ there will be a finite density of free vortices at length scales $\ell > \xi_+$, just above the transition this scale is still extremely large (see Fig.~\ref{fig:xi-nf}).
For qubit distances $z \ll \xi_+$, actually the dominant contribution still arises from the bound-pair contribution to Eq.~\ref{eqn:noise-epsilon}.
We note that $\Im \epsilon_v(\omega,q)\sim \omega$ as $\omega \to 0$ and therefore
\begin{multline}
    \label{eqn:eps-zero-w}
    \lim_{\omega \to 0} -\frac{1}{\omega}\Im \epsilon_v^{-1}(\omega,q) =\\
    \frac{1}{\epsilon^2_v(q)} \int_{\xi_c}^{\xi_+} \frac{d\ell}{\ell} \frac{\ell^2}{14 D} \left(\frac{d\epsilon_v}{d\log \ell}\right)F\left(q\ell/2\right).
\end{multline}
This can be integrated over $q$ to obtain the magnetic noise at low frequency which is 
\begin{multline}
    \lim_{\omega \to 0} \mathcal{N}_{zz}(\omega,z)/\mathcal{N}_0 = \\
   \frac{1}{14} \int_{0}^{\infty} dq qe^{-2zq}  \frac{\epsilon_c}{\epsilon^2_v(q)} \int_{\xi_c}^{\xi_+} d\ell \ell \left(\frac{d\epsilon_v}{d\log \ell}\right)F\left(q\ell/2\right).
\end{multline}
Although it depends on the particular details of the function $F$, if we use the Gaussian approximation (see Appendix~\ref{app:finite-q-eps}), one can show that within this approximation the static dielectric function obeys
\begin{equation}
    \frac{16}{q}\frac{d\epsilon^{-1}_v(q)}{dq} = \frac{1}{\epsilon_v^2(q)}\int_{\xi_c}^{\xi_+}d\ell \ell \frac{d\epsilon_v}{d\log \ell} F(q\ell/2).
\end{equation}
We can therefore show that~\footnote{It is again commented that this is not an exact relation due to the approximation used for $F(x)$, though we expect it to be a rather good approximation. In particular, it is unlikely that the numerical prefactor of $16/14$ is exact or universal.}
\begin{equation}
\label{eqn:noise-zero-w}
    \lim_{\omega \to 0} \mathcal{N}_{zz}(\omega,z)/\mathcal{N}_0 \approx \frac{16}{14}\epsilon_c \int_0^{\infty} dq e^{-2zq}\frac{d\epsilon_v^{-1}(q)}{dq}.
\end{equation}

\begin{figure}
    \centering
    \includegraphics[width=\linewidth]{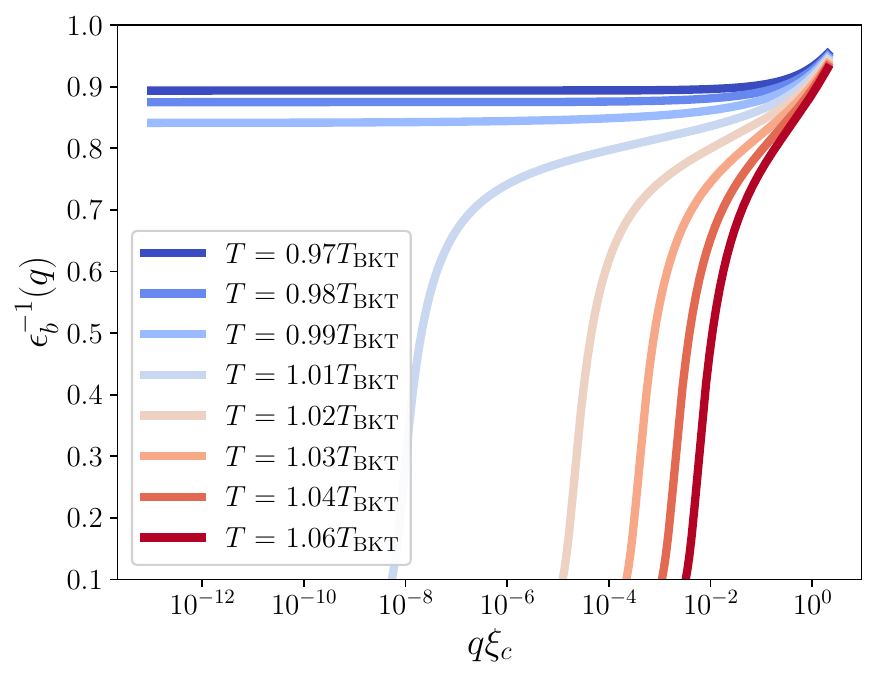}
    \caption{Inverse bound-vortex dielectric function $\epsilon_b^{-1}(q)$.
    For $T > T_{\rm BKT}$ the $\epsilon_b^{-1}(q)$ exhibits a crossover in $q$, vanishing at $q = 0$ and approaching unity as $q\to \xi_c^{-1}$.}
    \label{fig:eps_q}
\end{figure}

The integral is dominated by $q \lesssim 1/(2z)$ due to the exponential kernel, and therefore we can qualitatively expect it to behave as $\int_0^\infty dq e^{-2zq} \frac{d\epsilon_b^{-1}}{dq} \sim \int_0^{1/(2z)} dq \frac{d\epsilon_b^{-1}}{dq} = \epsilon_b^{-1}(q = 1/(2z)) - \epsilon_b^{-1}(0)$. 
For $T \gtrsim T_{\rm BKT}$ we have $\epsilon_b^{-1}(0) =0$ and therefore the magnetic noise will essentially follow the same crossover behavior as $\epsilon_b^{-1}(q)$. 
This is shown as a function of momentum $q$ for different temperatures $T$ in Fig.~\ref{fig:eps_q}, and closely parallels the behavior of $\epsilon_v^{-1}(\ell)$ in Fig.~\ref{fig:rg}(b). 
Conversely, at fixed $z$ (and hence $q\sim z^{-1}$) it is easy to see that as $T$ increases, the crossover length scale shrinks and ultimately becomes smaller than $2z$ at which point the noise rapidly drops to zero (in the presence of quasiparticles this will drop to a finite background). 
In Fig.~\ref{fig:noise-T} this rapid drop is observed for temperatures right above $T_{\rm BKT}$, but below the pseudo-plateau temperature set by $\xi_+(T)$. 

Physically, this local maximum in the noise can also be understood as a tradeoff between the dissipative and reactive parts of the vortex motion; at low temperatures the noise vanishes as the dissipation becomes frozen out in the superconducting phase.
At high temperatures, the magnetic noise also eventually drops since it originates from Johnson-Nyquist current fluctuations and therefore is proportional to the electrical conductance which decreases as superfluidity is destroyed.

This shows that by studying the dependence of the low frequency noise on the qubit-sample distance and temperature, one can probe the finite-size scaling behavior of the BKT transition directly. 
We now examine this in more detail below.  

\subsection{Distance Dependence}
\label{sub:dist}
\begin{figure*}
    \centering
    \includegraphics[width=\linewidth]{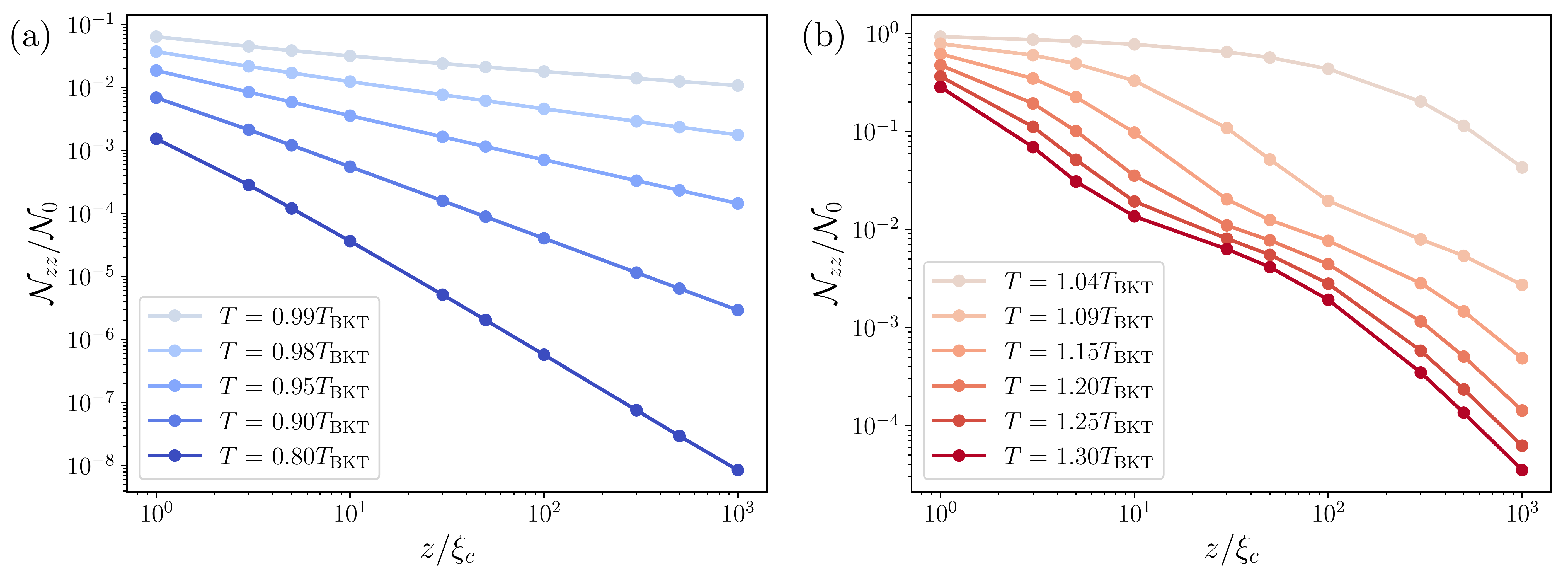}
    \caption{(a) Magnetic noise as $\omega \to 0$ for temperatures below the transition ($T < T_{\rm BKT}$).
    We see the dependence is essentially a power law which continuously varies with temperature, becoming progressively less sensitive to qubit distance $z$ upon approaching the transition. 
    (b) Magnetic noise at low frequency as $\omega \to 0$ for temperatures above the transition ($T > T_{\rm BKT}$).
    We see the dependence is more complicated, with at least two distinct regimes depending on whether the qubit distance $z$ is larger than the temperature dependent inter-vortex distance $\xi_+(T)$.  }
    \label{fig:noise-z}
\end{figure*}

We now study how the noise behaves in more detail as a function of the qubit-sample distance $z$.
This is presented as a function of $z$ in Fig.~\ref{fig:noise-z}(a),(b) for different temperatures below, and above the transition temperature, respectively.
Both are shown in the zero-frequency limit (we take $\omega = 10^{-25}\omega_0$ in our numerical calculations).

Below the BKT transition, the low-frequency scaling behavior seen in Fig.~\ref{fig:noise-z}(a) is relatively simple and exhibits a continuously-varying power law behavior.
This is expected and is consistent with known results~\cite{Ambegaokar.1978} since for $T < T_{\rm BKT}$ the system is generally characterized by algebraic correlations.

For $T > T_{\rm BKT}$ we see the behavior however is more complicated and is not a simple power law even as $\omega \to 0$.
To better understand this regime we focus on the contribution to the noise due to the free vortices (i.e. the vortex Drude weight).
This is expected to dominate for $z \gg \xi_+$ and low frequencies. 
If we neglect the frequency and momentum dependence of $\epsilon_b(q,\omega)$ we can write 
\begin{equation}
\label{eqn:epsilon-drude}
    \epsilon_v(\omega,q) = \epsilon_b\left[ 1 + \frac{\xi_{\rm D}^{-2}}{q^2 -i \ell_\omega^{-2}}\right],
\end{equation}
where $\epsilon_b = \epsilon_b(q=0,\omega=0)$ is the static renormalized dielectric contribution due only to the bound pairs, and 
\begin{equation}
    \xi_{\rm D}^{-2} = 4\pi^2 \frac{\rho_{\rm 2D}}{\epsilon_b} \frac{n_f}{T}
\end{equation}
is related to the Debye-Huckel screening length, while $\ell_{\omega}^{-2} = \omega/D$ characterizes the diffusion length for the vortices. 

In this regime one can find the zero-frequency limit of the magnetic noise is 
\begin{equation}
    \label{eqn:noise-zero-w-highT}
    \mathcal{N}_{zz}(0,z)/\mathcal{N}_0 = \frac{\epsilon_c}{14\epsilon_b}\int_0^\infty dq q e^{-2zq}\frac{\xi_{\rm D}^{-2}}{(q^2 + \xi_{\rm D}^{-2})^2}.
\end{equation}
In particular, this will exhibit the asymptotic behavior for $z \gg \xi_{\rm D}$ of 
\begin{equation}
    \mathcal{N}_{zz}(0,z)/\mathcal{N}_0 \sim \frac{\epsilon_c}{14\epsilon_b}\frac{1}{(2 z/ \xi_{\rm D} )^2}.
\end{equation}
This is what one would expect from local, scale-independent Johnson-Nyquist noise as predicted in Ref.~\onlinecite{Halperin.1979}, and this is reflected in the $z^{-2}$ power law tail in Fig~\ref{fig:noise-T}(b) at large $z$.

We have seen that there are interesting and detailed signatures of the BKT transition in the magnetic noise, which manifest in the distance dependence of the magnetic noise.
However, in general the frequencies probed by spin-qubit noise magnetometry are not negligibly low, and this may introduce additional complications into the predictions, as well as potentially offer novel insights. 
It is therefore important to understand the dependence of the noise also as a function of frequency in order to get a more complete picture.

\subsection{Frequency Dependence}
\label{sub:freq}
\begin{figure}
    \centering
    \includegraphics[width=\linewidth]{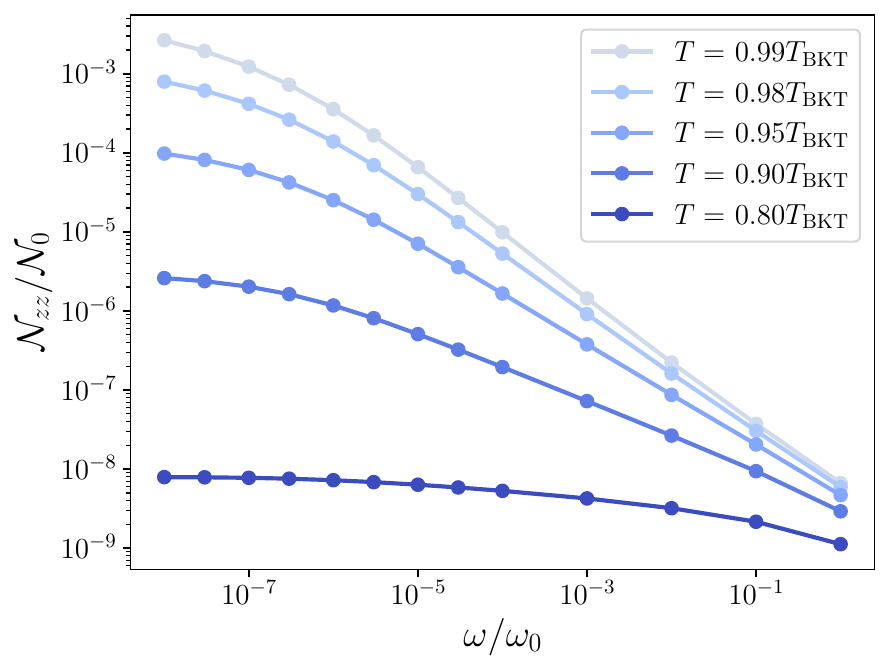}
    \caption{Magnetic noise at $z = 1000\xi_c$ as a function of $\omega$ for different $T < T_{\rm BKT}$.
    We see two regimes. 
    At very low frequencies the noise is essentially white noise and independent of frequency.
    Then, at higher frequencies we see a crossover to power law behavior with a varying exponent. 
    }
    \label{fig:noise-w-lowT}
\end{figure}

We now study the full frequency dependence of the magnetic noise for different temperatures and probe depths.
We start by analyzing the magnetic noise at low temperatures, below and at the effective BKT transition as a function of frequency.
This is shown in Fig.~\ref{fig:noise-w-lowT} for a variety of temperatures below $T_{\rm BKT}$ for $z = 1000\xi_c$, essentially probing the long-wavelength magnetic noise. 

\begin{figure}
    \centering
    \includegraphics[width=\linewidth]{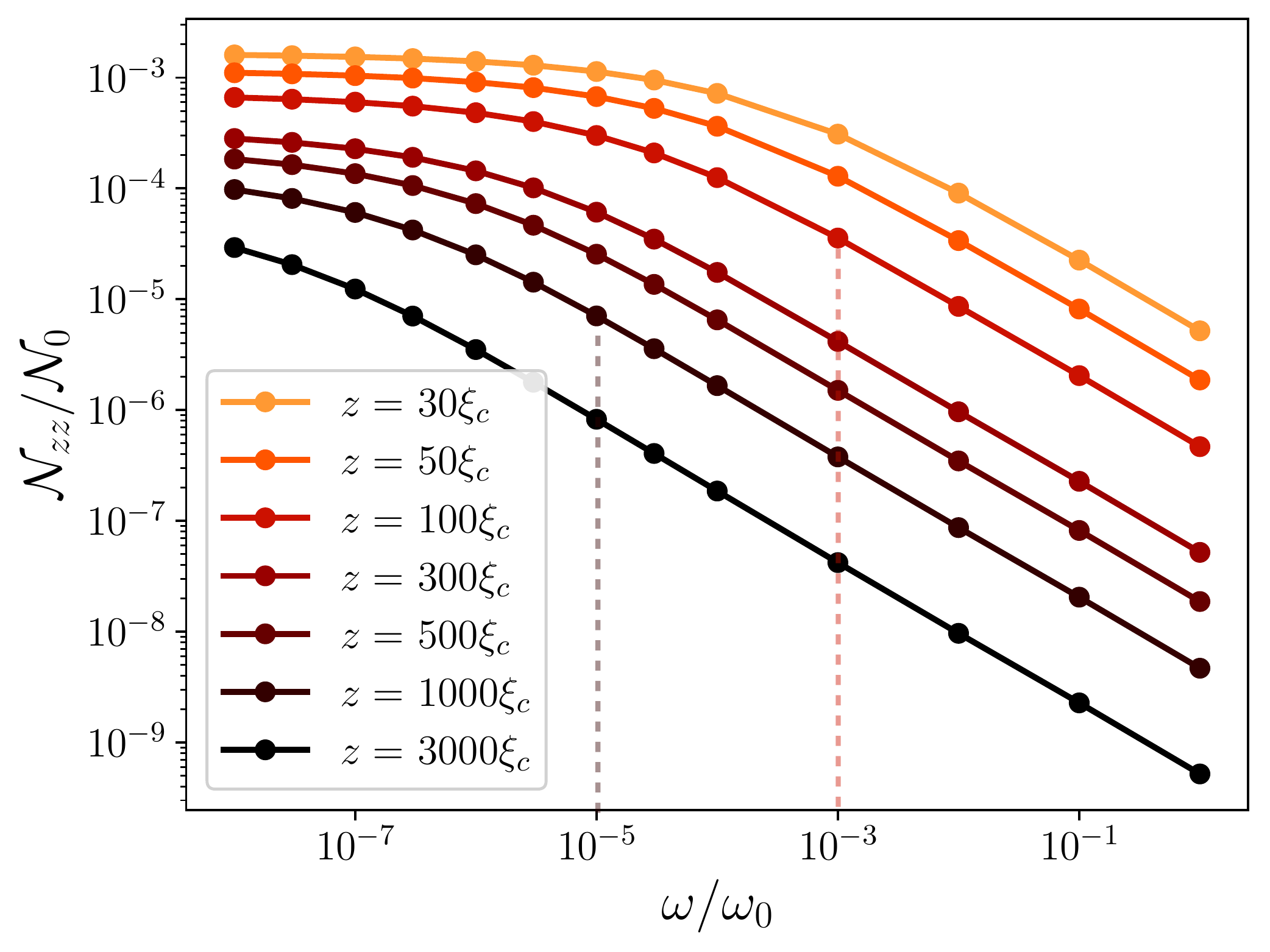}
    \caption{Magnetic noise at $T = 0.95T_{\rm BKT}$ as a function of $\omega$ for different $z$ values.
    Dashed vertical lines are guides to the eye showing where the crossover frequency $\omega_z$ to power law behavior occurs.
    Increasing $z$ by a factor of 10 reduces the crossover frequency by a factor of 100, confirming the scaling exponent of $\omega_z \sim z^{-2}$.
    }
    \label{fig:noise-w-zs-lowT}
\end{figure}

Looking more closely we see that there are two clear noise regimes. 
At very low frequencies, the noise is approximately white noise with no strong frequency dependence, thereby justifying our zero-frequency analysis of Eq.~\eqref{eqn:noise-zero-w}.
This persists up until a $z$-dependent frequency $\omega_z$, above which the noise character changes to a power law in frequency with an exponent that continuously varies with temperature.
This again is similar to our observations in Fig.~\ref{fig:noise-z}(a) and essentially recovers the results of previously performed bulk flux-noise measurements~\cite{Kim.1999,Korshunov.2002,Minnhagen.1987,Houlrik.1994,Shaw.1996,Bjornander.1996,Festin.1999,Rogers.1992}.
In Fig.~\ref{fig:noise-w-zs-lowT} we clearly see this crossover behavior and confirm the scaling of $\omega_z \sim z^{-2}$ as expected based on diffusion.

For frequencies $\omega \gg \omega_z$ (i.e. much larger than the crossover scale identified) we can analytically understand the power law behavior in frequency.
In this case, the noise can be written as 
\begin{equation}
    \lim_{z \to \infty} \mathcal{N}_{zz}(\omega,z)/\mathcal{N}_0 \approx -\frac{\pi}{4}\frac{D\epsilon_c}{\omega (2z)^2} \frac{d\epsilon_v^{-1}(\ell)}{d\log \ell}\bigg|_{\ell = \sqrt{14 D/\omega}}.
\end{equation}
This exhibits a leading dependence on distance as $z^{-2}$, as one would expect based on magnetostatics from local fluctuations of the sheet-current density.
The fact that this also exhibits a power law in frequency (at fixed $z$) is less trivial and originates from the anomalous scaling behavior of the vortex dielectric function $\epsilon_v^{-1}(\ell)$.
This behavior has been previously studied in Ref.~\onlinecite{Ambegaokar.1978} in the $\mathbf{q} \to 0$ limit and can be related to the scaling exponent $x(T)$ which describes the power law correlations in the QLRO phase. 
Specifically, we expect 
\begin{equation}
    \mathcal{N}_{zz}(\omega,z)/\mathcal{N}_0 \sim z^{-2} \omega^{x(T)/2 -1}.
\end{equation}
The exponent $x(T)$ continously varies with temperature and near $T_{\rm BKT}$ goes as 
\begin{equation}
    x(T)/2 \propto \sqrt{1 - T/T_{\rm BKT}},
\end{equation}
with a nonuniversal constant as the prefactor~\cite{Ambegaokar.1978}.
Since this vanishes as $T\to T_{\rm BKT}$ we see that the noise tends to $1/\omega$ behavior over a large range of frequency close to the transition, also reproducing previously known results~\cite{Korshunov.2002}.

Finally, at very high frequencies with $\omega \gtrsim \omega_0$ the continually varying power law behavior gives way to a $1/\omega^2$ behavior (this is not shown in the figures for brevity).
This is seen by performing a high-frequency expansion of Eq.~\eqref{eqn:epsilon_wz}.

\begin{figure}
    \centering
    \includegraphics[width=\linewidth]{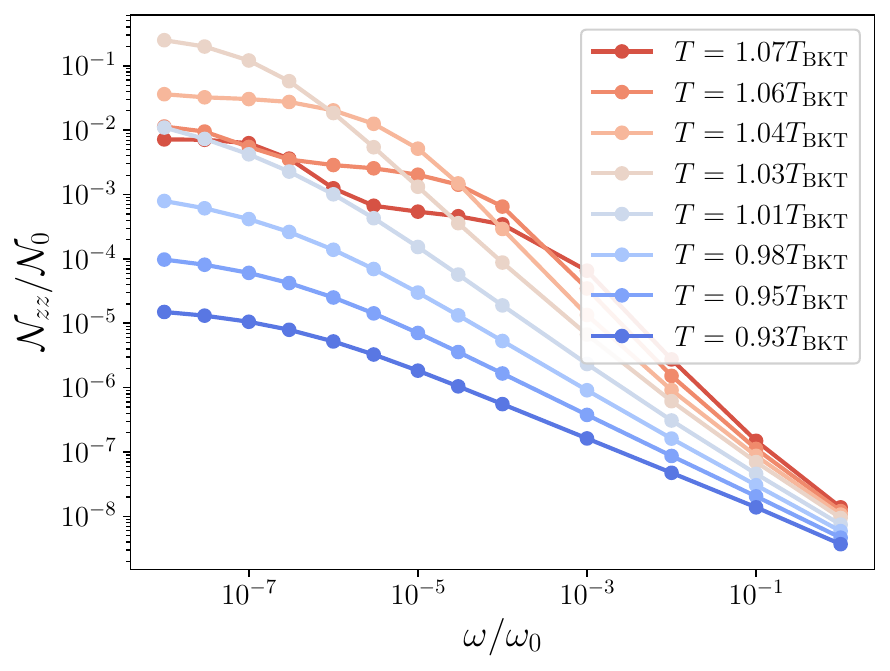}
    \caption{Magnetic noise at $z = 1000\xi_c$ as a function of $\omega$ for different $T \sim T_{\rm BKT}$. 
    At low temperatures we see the crossover from white noise to a power law, but passing through the transition we note a number of new features including the emergence of a second plateau in the noise and nonmonotonic temperature dependence.}
    \label{fig:noise-w-BKT}
\end{figure}

We now turn out attention to understanding the behavior as $T$ varies through the BKT transition.
This is seen in Fig.~\ref{fig:noise-w-BKT}.
We first point out that at low frequencies the magnetic noise is nonmonotonic with temperature, initially increasing as the transition is approached from below before dropping again once the temperature surpasses $T_{\rm BKT}$, as seen also in Fig.~\ref{fig:noise-T}, which is essentially the $\omega \to 0$ extrapolation of this dependence. 

We also see that for $T>T_{\rm BKT}$ a new frequency scale emerges which again reflects the appearance of the vortex Drude weight (this is only visible provided $T$ is sufficiently large that $\xi_+$ is comparable to the relevant length scale $z$).
The free vortex contribution to $\epsilon$ is modeled in Eq.~\eqref{eqn:epsilon-drude}; rewriting in terms of frequency scale we have 
\begin{equation}
    \epsilon_v(\omega,q) = \epsilon_b + \frac{\gamma}{\omega_q - i\omega}
\end{equation}
where $\omega_q = Dq^2$ and 
\begin{equation}
    \gamma = 4\pi^2 \rho_{\rm 2D}n_f \mu 
\end{equation}
is the Drude weight of the free vortices. 
Note $\gamma/\omega_0 = 8\epsilon_c(\xi_+/\xi_c)^{-2} $, and thus $\gamma$ may be many orders of magnitude smaller than $\omega_0$ due to the strong dependence of $\xi_+$ on temperature.
Below this frequency the magnetic noise is well reproduced by Eq.~\eqref{eqn:noise-zero-w-highT}.

For frequencies above $\gamma$ there is then another apparent plateau which emerges in the magnetic noise, clearly seen on the highest temperature curves in Fig.~\ref{fig:noise-w-BKT}.
In this regime the free vortices are no longer resolved for $\omega \gg \gamma$, and instead we see the remnant of the magnetic noise due to the bound-pair contribution to $\epsilon_v$ which are still important even for $T>T_{\rm BKT}$.
This then persists up until a second crossover scale before the noise again drops at higher frequencies, and ultimately recovers the $1/\omega^2$ behavior for $\omega \sim \omega_0$.

\begin{figure*}
    \centering
    \includegraphics[width=\linewidth]{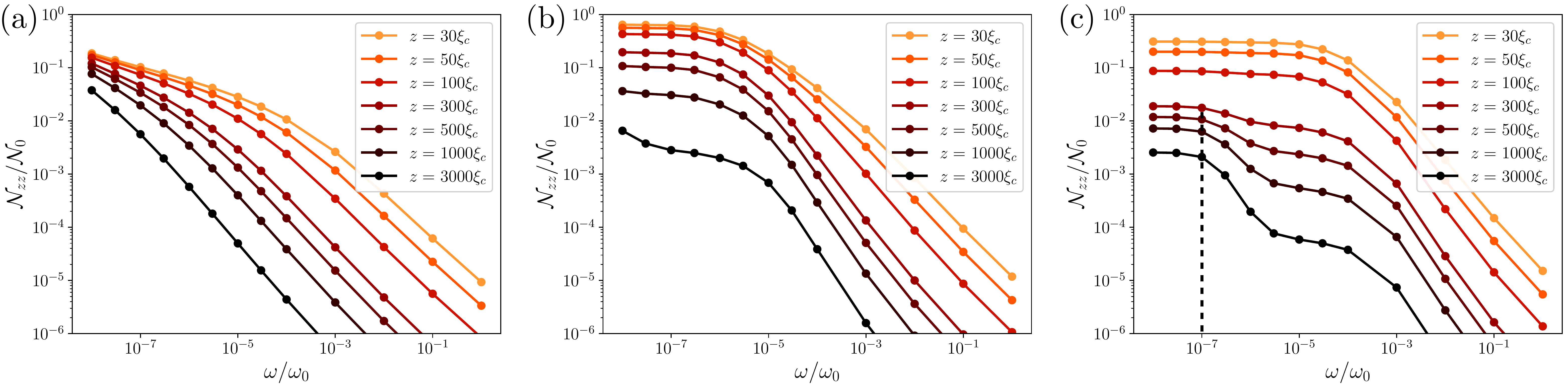}
    \caption{(a) Magnetic noise at $T = 1.02T_{\rm BKT}$ for different $z$ as a function of $\omega$.
    (b) Magnetic noise at $T = 1.04T_{\rm BKT}$ for different $z$ as a function of $\omega$.
    We see the emergence of white noise for $\omega <\gamma $, which grows out from low-frequencies while also flattening out in $\omega$.
    (c) Magnetic noise at $T = 1.07T_{\rm BKT}$ for different $z$ as a function of $\omega$.
    The white noise regime is clearly defined and extends up to higher frequencies.
    We also see the emergence of the Drude weight for frequencies below the $z$-independent dashed line. 
    }
    \label{fig:noise-w-zs-highT}
\end{figure*}

We now focus on understanding the $\omega$ and $z$ dependence of the magnetic noise in the high-temperature regime, which is summarized in Fig.~\ref{fig:noise-w-zs-highT}(a),(b),(c) for temperatures $T/T_{\rm BKT} = 1.02,\ 1.04,\ 1.07$ respectively. 
For temperatures close to the transition, the dependence still closely resembles the $T < T_{\rm BKT}$ dependence at the relevant frequency and length scales, although there is convergence of the curves for low-frequencies as a function of $z$ owing to the proximity to the scaling collapse which is expected at the transition itself. 

Progressively increasing the temperature causes the magnetic noise to grow and flatten out at low frequencies as the crossover scale in the bound-pair contribution grows larger.
For the longest length scales, signatures of the vortex Drude weight also become visible, as seen in the $z=3000\xi_c$ curve in Fig.~\ref{fig:noise-w-zs-highT}(b), though the contribution is hardly visible.

Finally, at even higher temperatures we see the emergence of the vortex Drude weight, indicated by the dashed line which is a visual guide, roughly corresponding to frequency $\omega = \gamma$.
Below this frequency, the dominant noise source are the free vortices and their diffusion, while above this frequency, the bound-pairs are the main source of noise.
We clearly can see here that this crossover frequency is not dependent on distance $z$ but only on temperature $T$, which heralds the intrinsic length scale $\xi_+(T)$ appearing in the problem.

The magnitude of the drop in noise at this frequency is however $z$ dependent, with larger length scales exhibiting a more dramatic difference between the free and bound pairs.
This is simply understood as the fact that at larger length scales the bound pairs become even less important and the contribution to magnetic noise from these bound pairs drops more rapidly as $z\to \infty$ than the contribution from free vortices does, which only drops as $z^{-2}$.

This in-depth analysis shows that the joint frequency, distance, and temperature dependence of the magnetic noise can be used to disentangle and understand the behavior of both the bound and free vortices and their dynamics above as well as below the BKT transition.
All of this very elegantly follows from the analysis of the superconducting phase fluctuations in the London limit, which assumes that the amplitude of the superconducting order is frozen and the only fluctuations are those of the phase (except possibly at the vortex core).
However, given the practical nature of the BKT transition in superconductors, which are actually formed from Cooper pairs that are extended objects formed from paired fermions, it is worthwhile to compare these results to those that might be expected based on a simpler analysis of Gaussian superconducting fluctuations. 
We therefore briefly investigate this aspect in the next section.

\section{Aslamazov-Larkin Fluctuations}
\label{sec:AL}
Up to this point, all of our calculations have been within the framework of the London limit, which realizes the XY model for length scales beyond the coherence length $\xi_c$, and is built upon the assumption that amplitude fluctuations of the pair condensate, as well as electronic quasiparticle excitations have been frozen out at low temperatures.
This is however not realistic for many cases where $T_{\rm BKT} \sim T_{\rm BCS}$, the critical temperature within the BCS framework and, more generally the temperature at which pairing sets in.

In this case, in addition to vortex fluctuations, Gaussian fluctuations of the order parameter $\psi(\mathbf{r},t)$ are also important and may also contribute to the transverse electromagnetic noise that decoheres the spin qubit.
Here we will briefly compare these fluctuations to the non-Gaussian vortex fluctuations considered in the primary part of this paper and argue that they may be distinguished from each other based on the behavior of the qubit noise.
In particular, we identify a qualitative difference in the magnetic noise as a function of distance $z$ which allow one to discriminate between these two models of superconducting fluctuations. 

We specifically consider order parameter fluctuations within a time-dependent Ginzburg-Landau framework, similar to Ref.~\onlinecite{Halperin.1979,Mikeska.1970} and found in great detail in, e.g. Ref.~\onlinecite{Larkin.2005}.
In order to capture the dynamical noise we use a Langevin model for the order parameter dynamics of the form 
\begin{equation}
\label{eqn:tdgl}
     \Gamma^{-1} \partial_t \psi(x) = -\frac{\delta \mathcal{F}}{\delta \psi^*(x)} +\eta(x)
\end{equation}
where $\Gamma^{-1} = \nu_F \tau_{\rm GL} $ is the kinetic coefficient for relaxation, $\eta(x)$ is a complex noise field which satisfies a fluctuation dissipation relation, and the free energy is given (near $T_{\rm BCS}$) by  
\begin{equation}
\label{eqn:glfe}
     \mathcal{F} = \nu_F \int d^2 r \left[ \tilde{\xi}_{\rm BCS}^2 |\nabla \psi|^2 + r |\psi|^2 + \frac12 u |\psi|^4 \right]
\end{equation}
where $\tilde{\xi}_{\rm BCS}$ is a length scale roughly corresponding to the BCS coherence length at low temperatures, $\nu_F$ is the density-of-states at the Fermi level and $r = (T - T_{\rm BCS})/T_{\rm BCS}$. 
$u>0$ is a parameter which reflects the nonlinearity of the condensate.
In this work we will focus on the thermal current fluctuation spectrum at low frequencies as a function of momentum $\mathbf{q}$ for $T > T_{\rm BCS}$.
In this case, we will drop the nonlinear term $u|\psi|^4$ and restrict to $T$ not too close to $T_c$, though in principle this can be relaxed using a self-consistent Gaussian approximation (e.g. as done in Ref.~\cite{Curtis.2023} for the case of a spinful order parameter). 
We also neglect other contributions which may arise from the order parameter fluctuations, including the so-called ``density-of-states" and ``Maki-Thompson" contributions~\cite{Larkin.2005}, leaving these to future works. 

First, we compute the sheet current density.
This has contributions from the diamagnetic response of the pairs as well as the paramagnetic response.
In the absence of an applied field, to quadratic order the diamagnetic term does not contribute, and furthermore is non-dissipative and will not show up in the noise response.
Therefore, we can focus on the paramagnetic response only which, in momentum space is given by 
\begin{equation}
\label{eqn:AL-current}
    \mathbf{j}_\square(\mathbf{q},t) = 2(2e) \nu_F \tilde{\xi}_{\rm BCS}^2 \int_{\bf p} \mathbf{p} \overline{\psi}_{\mathbf{p}+\mathbf{q}/2}(t){\psi}_{\mathbf{p}-\mathbf{q}/2}(t).
\end{equation}
From this we can compute the current noise spectral function $S^\perp(\mathbf{q},\omega)$ which is the object of importance. 
For details we refer the reader to Appendix~\ref{app:AL}.
Then, the current fluctuations 
\begin{equation}
    S_{ab}(\mathbf{q},\omega) = \int dt e^{i\omega t} \langle j^a_\square(\mathbf{q},t) j^b_\square(-\mathbf{q},0) \rangle
\end{equation}
can be computed by taking advantage of the fact that the fluctuations are Gaussian. 

At $\omega = 0$ we find the magnetic noise from Gaussian pair fluctuations is 
\begin{multline}
    \mathcal{N}_{zz}/\mathcal{N}_0^{\rm AL} = (1+r)^2 \\
    \times \int_0^\infty dx \frac{e^{-4x z/\tilde{\xi}_{\rm BCS}}}{x}\int_0^\infty du u \\
    \times \left[\frac{  (x^2 + u^2 + r) - \sqrt{\left[ (u+x)^2 +r \right]\left[ (u-x)^2 +r\right]}}{(u^2 + x^2 +r)^2} \right].
\end{multline}
The normalization constant is $\mathcal{N}_0^{\rm AL} = (e\mu_0/\pi)^2 T_{\rm BCS}^2\tau_{\rm GL}/\tilde{\xi}_{\rm BCS}^2$, and we have expressed this as a function of $r >0 $.

\begin{figure*}
    \centering
    \includegraphics[width=\linewidth]{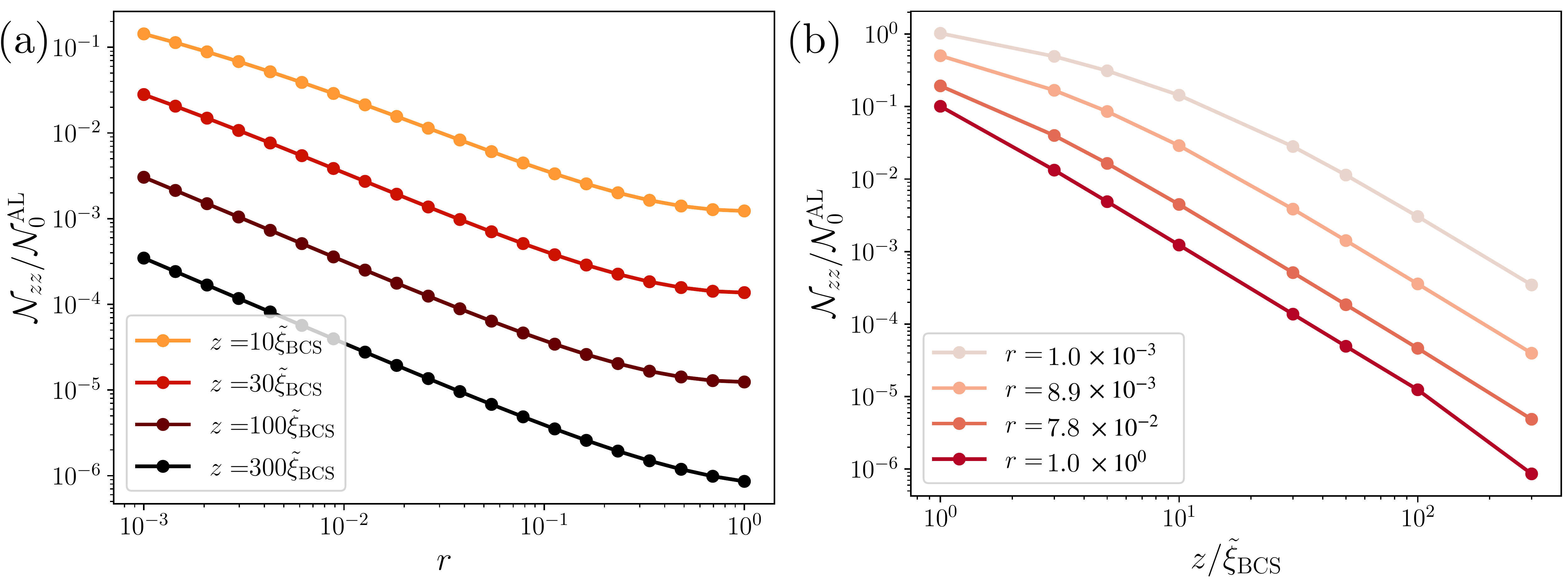}
    \caption{(a) Local magnetic noise at $\omega = 0$ as a function of $r = T/T_{\rm BCS} - 1 >0$ for different distances $z$ due to Aslamazov-Larkin Gaussian superconducting fluctuations.
    (b) Local magnetic noise at $\omega = 0$ as a function of distance $z$ for different $r>0$.}
    \label{fig:noise-AL}
\end{figure*}

The magnetic noise is depicted in Fig.~\ref{fig:noise-AL} for temperatures above, but near the transition $T_c$ (note here we indicate the temperature that superconductivity sets in as $T_c$ since it is not a BKT transition in this case, but it should be regarded as equivalent).
In Fig.~\ref{fig:noise-AL}(a) we present this as a function of temperature (technically $r$) for different qubit-sample distances $z$, which is to be compared to Fig.~\ref{fig:noise-T} (though note the different $x$-axis scales and ranges).
In Fig.~\ref{fig:noise-AL}(b) we present the same data but as a function of distance $z$ for different temperatures $T$, which is to be compared to Fig.~\ref{fig:noise-z}(b). 

From the analytic formula, evaluated at $r = 0\Rightarrow T = T_{\rm BCS}$, we find that the low-frequency magnetic noise is also varies as $\log(z/\ell_{\rm max})$ at the transition point, with an infrared cutoff $\ell_{\rm max} = \textrm{min}(L,\sqrt{D/\omega})$ which is taken to be the minimum of the sample size $L$ or the diffusive length $\ell_{\omega} = \sqrt{D/\omega}$.
Explicitly, we find 
\begin{equation}
    \lim_{\omega\to 0,T\to T_{\rm BCS}}\mathcal{N}_{zz}/\mathcal{N}_0^{\rm AL} = \log\left(2\right)\times \textrm{Ei}_1(4z/\ell_{\rm max}).
\end{equation}
Here $\textrm{Ei}_1(u) = \int_u^{\infty}e^{-s}/s ds$ is the exponential integral function, which diverges as $z \to 0$ like $\log(z)$. 
However, we will emphasize that this logarithmic dependence is only expected to manifest very close to the critical point, where we anyways expect our non-interacting Gaussian approximation used here to break down.
Further from the critical point it can be seen that the overall scaling of the noise at a distance $z$ is dominated at large $z$ by magnetostatic scaling from the standard Aslamazov-Larkin paraconductivity as 
\begin{equation}
    \mathcal{N}_{zz}/\mathcal{N}_0^{\rm AL} \sim z^{-2}/(T - T_c).
\end{equation}
This is the power law we are able to see in Fig.~\ref{fig:noise-AL}, which again is only expected to hold for $T$ not too close to $T_c$.

In order to distinguish between the Aslamazov-Larkin Gaussian fluctuations and the BKT fluctuations, we turn to studying the distance dependence for $T > T_{\rm BKT}$.
In this case there is a clear feature in the distance dependence in Fig.~\ref{fig:noise-z}(b) which can be resolved due to the vortex Drude weight and associated density $n_f$, in addition to the crossover scale that is due to the bound-pairs. 
In contrast, the distance dependence in Fig.~\ref{fig:noise-AL}(b) only has a single crossover scale, governed by the correlation length of the superconducting fluctuations.
The qualitative difference in the distance dependence and in particular the contribution of bound-pairs for smaller probe distances in the case of the BKT physics is a key signature by which one may discriminate between the Gaussian fluctuations and the vortex unbinding fluctuations. 
While we leave a more detailed study to future works, we can already see that the ability to probe the magnetic noise as a function of depth in the vicinity of the transition may be extremely helpful in disentangling the Gaussian Aslamazov-Larkin fluctuations from the topological vortex fluctuations which characterize the BKT transition.

Armed with this knowledge, we will now turn to a survey of potential materials which may realize the BKT transition in suitable parameter regimes and outline the strengths and constraints of various candidate systems. 

\section{Experimental Considerations}
\label{sec:expt}

In the preceding sections we have theoretically identified a number of signatures of the BKT transition which are expected to manifest in the local magnetic noise spectrum.
Before providing experimental estimates, it is important to reiterate some of the physical assumptions of our calculations.
In particular, we have assumed (i) that fermionic quasiparticles are sufficiently gapped out and don't contribute to the magnetic noise; this will be briefly revisited later in Sec.~\ref{sub:qp}.
We have also assumed (ii) that the superfluid density and coherence length don't exhibit appreciable temperature dependence and can be treated as constants within our model.
Finally, we have assumed (iii) a particular model for the bare vortex fugacity of $y_0(T) = 0.1 \exp\left( -T_{\rm BKT}^{(0)}/T\right)$, which may also lead to some model dependence of our quantitative results.
Nevertheless, it is worthwhile to consider a few possible material systems which may realize our theoretical predictions.

First, to assess the potential viability of NV noise magnetometry for measuring the phenomena described in the present work, we estimate the relevant fundamental scales which set the parameters of the problem.
These are the coherence length $\xi_{c}$, the frequency $\omega_0 = 2\pi \nu_0 = D_0\xi_c^{-2}$ (which in turn is set by the vortex mobility), and the magnetic noise scale $\mathcal{N}_{0}$ (Eq.~\ref{eq:noise_scale}). The following expression is used to estimate the vortex mobility, applicable to dirty superconductors, following the Bardeen-Stephens model~\cite{Halperin.1979,Bardeen.1965}
\begin{equation}
    \mu = 4\xi_{c}^{2}\frac{R_{\square}}{R_{K}},
\end{equation} where $R_\square$ is the normal state resistance per square and $R_K  = h/e^2 \sim 25.8\si{\kilo\ohm}$ is the resistance quantum, and $\hbar = k_{B} = 1$.

Our estimates, listed in Table~\ref{table:materials}, are made using the measured $T_{\rm BKT}$. Coherence lengths in the literature reported for these materials are determined based on the upper critical field (see references in Table~\ref{table:materials}). In the case of MoS$_2$ and FeSe on SrTiO$_3$, the coherence length was determined on a different sample from the one where superfluid density was measured. For 1 UC Bi$_2$Sr$_2$CaCu$_2$O$_{8+\delta}$, a representative coherence length was obtained from measurements on the bulk material~\cite{naughton1988orientational}.

Additionally, it is useful to compare the BKT transition temperature, $T_{\rm BKT}$, and the mean-field superconducting transition temperature ($T_{\rm BCS}$) where it is known. The separation between $T_{\rm BKT}$ and $T_{\rm BCS}$ determines the range of temperatures where the present theory applies.


\begin{table*}[t]
\begin{center}
\begin{tabular}{|c | c  c c c | c c c|} 
 \hline
 Material &  $T_{\rm BKT}$\ (\text{K}) & $T_{\rm BCS}$ (K) & $\xi_{c}$\ (\text{nm}) & $R_{\Box}$ ($\Omega$) & $\nu_0$\ \textrm{(GHz)}& $\mathcal{N}_{0}$ (\textrm{pT}$^{2}$/\textrm{Hz}) & $\mathcal{N}_{0}$ (Hz)\\ [0.5ex] 
 \hline\hline
 1 UC FeSe on SrTiO$_{3}$ ~\cite{Zhao.2016,biswas2018direct} & 20.3 & - & 2 - 3 & 500 - 600 & 33 - 40 & 4000 - 11000 & 3 - 9 \\ 
    \hline
 1 UC Bi$_{2}$Sr$_{2}$CaCu$_{2}$O$_{8+\delta}$~\cite{Yu.2019,naughton1988orientational} &  84.2 & - & 3.1$^{*}$ & 1000 & 270 & 10000 & 8\\
    \hline
gated MoS$_{2}$ film~\cite{jarjour2023superfluid,saito2016superconductivity} &  9.5 & - & 8 & 750 & 23 & 220 & 0.2\\
 \hline
  3.5 nm NbN film~\cite{weitzel2023sharpness}  & 4.5 & 5.75 & 6.7 & 4100 & 100 & 45 & 0.04\\
 \hline
MATTG~\cite{park2021tunable}  & 2.3 & - & 12.5 & 3000 & 22& 5.4 & 0.0042\\
 \hline
\end{tabular}
\end{center}
 \caption{Estimate of scales relevant to two-dimensional superconductors. 
 The frequency scale $\nu_0 = \omega_0/(2\pi)$ is typically many orders of magnitude larger than the relevant experimental frequencies of the qubit probe (see Sec.~\ref{sub:freq}).
 Magnetic noise is presented in both pT$^2$/Hz and an equivalent NV relaxation rate in Hz, which is estimated by multiplying the magnetic noise by the NV center's gyromagnetic ratio squared (0.028 Hz/pT)$^{2}$. 
 $^{*}$representative coherence length obtained from the bulk material.}
 \label{table:materials}
\end{table*}

Many of the superconductors listed in Table~\ref{table:materials} are predicted to exhibit magnetic noise intensities, $\mathcal{N}_{0}$, that are within the sensitivity of state-of-the-art $T_{1}$ relaxometry experiments. For example, magnetic field noise intensities of 10s of pT$^{2}$/Hz have been detected using $T_{1}$ relaxometry~\cite{andersen2019electron}. In practice the NV-sample distance can be realistically varied in such an experiment between 10 nm to 100s of nm~\cite{kolkowitz2015probing}. For the 2D superconductors listed in Table~\ref{table:materials}, such a distance-dependent relaxometry experiment would cover a broad range of length scales relevant to vortex diffusion (Fig.~\ref{fig:noise-z}) from distances on the order of $\xi_{c}$ up to an order of magnitude larger. The frequency bandwidth of $T_{1}$ relaxometry is set by the NV level splitting, which for the $|0\rangle \rightarrow|-1\rangle$ transition, can be tuned continuously from 2.87 GHz to 0 GHz by varying an external magnetic field from 0-100 mT along the NV axis. Thus, exploration of frequency scaling of the magnetic noise (Figs.~\ref{fig:noise-w-lowT},\ref{fig:noise-w-zs-lowT},\ref{fig:noise-w-BKT},\ref{fig:noise-w-zs-highT}) would also be possible using $T_{1}$ relaxometry as long as these relatively small applied fields do not significantly affect BKT vortex dynamics.
For a two-dimensional spin-singlet superconductor this is a relatively safe requirement. 
For sufficiently thin films the in-plane fields are limited by the Clogston-Chandrasekhar limit of $H_{c}/T_{\rm BCS} = 1.8 \;\si{\tesla\per\kelvin}$~\cite{Clogston.1962,Chandrasekhar.1962} and thus for all systems considered here we are well below the critical field since $T_c \gtrsim 1\;\si{\kelvin}$.

For many of the superconductors considered in Table~\ref{table:materials}, the proposed experiment would be challenging but viable. Monolayer high-temperature superconductors are perhaps the most promising in terms of absolute magnetic noise intensity. NbN, however, has the advantage of having a known separation between the mean-field and BKT transition temperatures, as well as ease of fabrication. In this case, the signal could be amplified by layering NbN films with a thin insulator between them.

We now briefly assess the background magnetic noise due to Bogoliubov quasiparticles, which is expected to be one of the largest sources of background noises that needs to be overcome in order to observe the BKT physics successfully. 

\subsection{Quasiparticle Effects}
\label{sub:qp}
Here we briefly discuss the impact of residual Bogoliubov-de Gennes (BdG) quasiparticles and their contribution to the magnetic noise near $T_{\rm BKT}$.
A detailed analysis of this has already been carried out and can be found in Ref.~\cite{Dolgirev.2022,Chatterjee.2022}, however we will discuss this here for completeness.

In particular, the principle result of Fig.~\ref{fig:noise-T} is presented assuming there are no residual BdG quasiparticles.
In cases where the BKT transition is clearly visible this assumption is likely valid since if $T_{\rm BKT} \ll T_{\rm BCS}$ there should already be an appreciable spectral gap.
However, especially in the case of strong disorder or nodal pairings this assumption can be called in to question. 

In order to address this, we will consider the impact on the transverse conductivity, the real part of which is responsible for magnetic noise.
In Appendix~\ref{app:conductivity} we show that the vortex contribution to the frequency and momentum dependent conductivity is  
\begin{equation}
    \sigma^\perp_v(\omega,{q}) = -\frac{ \rho_{\rm 2D}}{i\omega \epsilon_v(\omega,{q})}.
\end{equation}
If we take the quasiparticle conductivity to contribute as a momentum independent background (assumed valid for length scales longer than the quasiparticle mean-free-path) we can model the total conductivity $\sigma^\perp = \sigma^\perp_v + \sigma^\perp_{\rm qp}$ as 
\begin{equation}
    \sigma^\perp(\omega,{q}) = -\frac{ \rho_{\rm 2D}}{i\omega \epsilon_v(\omega,{q})} + \sigma_{\rm qp}(\omega).
\end{equation}
Here $\sigma_{\rm qp}$ is the frequency-dependent normal fluid conductivity due to residual quasiparticles.
In the case of a dirty $s$-wave conventional superconductor this is found from the standard Mattis-Bardeen result~\cite{Mattis.1958}.

It is easiest to distinguish the effects of vortices at low frequencies.
We therefore aim to understand the behavior of $\sigma_{\rm qp}$ at low frequency.
For $T>T_{\rm BCS}$ we will use $\sigma_{\rm qp}(\omega) = \sigma_n$, where $\sigma_n$ is the normal-state conductivity, assuming frequency dependence is captured effectively by $\omega \to 0$ limit in this regime.
For $T<T_{\rm BCS}$ we must keep the frequency dependence.
From Ref.~\onlinecite{Mattis.1958} we find that
\begin{multline}
   \textrm{Re} \left[ \sigma_{\rm qp}(\omega)/\sigma_n \right] = \\
 2 \int_{\Delta}^\infty  \frac{dE}{\omega} \left[f(E)-f(E+\omega)\right]\left[ \frac{E(E+\omega)+\Delta^2}{\sqrt{E^2 - \Delta^2}\sqrt{(E+\omega)^2 -\Delta^2}}\right] \\
 - \theta(\omega>2\Delta)\int_\Delta^{\omega} \frac{dE}{\omega}\tanh(\frac{E}{2T})\left[ \frac{E(E-\omega)+\Delta^2}{\sqrt{E^2 - \Delta^2}\sqrt{(E-\omega)^2 -\Delta^2}}\right] .
\end{multline}
In this expression, there is also a temperature dependence of $\Delta(T)$ which we capture using a BCS interpolation formula 
\begin{equation}
    \Delta(T)/T_{\rm BCS} = 1.76\tanh(1.74\sqrt{T_{\rm BCS}/T - 1}).
\end{equation}

Using this conductivity we can compute the magnetic flux noise including both the vortex and quasiparticle channels as 
\begin{equation}
    \mathcal{N}_{zz}^{\rm tot}  = \frac12 T \mu_0^2\int_{\bf q} e^{-2zq}\textrm{Re}\left[ -\frac{ \rho_{\rm 2D}}{i\omega \epsilon_v(\omega,{q})} + \sigma_{\rm qp}(\omega) \right].
\end{equation}
This involves the vortex-generated magnetic noise derived in Eq.~\eqref{eqn:noise-epsilon}, as well as a background term $\mathcal{N}_{zz}^{\rm qp}$ which, when normalized by $\mathcal{N}_0$ gives 
\begin{equation}
    \mathcal{N}_{zz}^{\rm qp}/\mathcal{N}_0  = \frac{\pi^2}{4}  \frac{T}{T_{\rm BKT} } \mu \int_{\bf q} e^{-2zq}\textrm{Re}\left[  \sigma_{\rm qp}(\omega) \right].
\end{equation}
We recall that, within the Bardeen-Stephens model, $\mu = 4 \xi_c^2 /\sigma_n$ using $\sigma_n = 1/R_\square$ in natural units. 
Crucially, we expect that the quasiparticle conductivity will largely be local for the relevant length scales pertinent to vortex dynamics.
Therefore we expect a quasiparticle background contribution to the noise (normalized to the same scale as the vortex signal), which scales as $z^{-2}$ for all temperatures as 
\begin{equation}
\label{eqn:qp-noise}
    \mathcal{N}_{zz}^{\rm qp}/\mathcal{N}_0  = \frac{\pi}{2} \frac{T}{T_{\rm BKT} } \frac{\xi_c^2}{4z^2}  \textrm{Re}\left[  \sigma_{\rm qp}(\omega)/\sigma_n \right].
\end{equation}

In order to make a direct connection with the vortex noise, we need an estimate for the ratio of $T_{\rm BCS}$ to $T_{\rm BKT}$, which controls the quasiparticle gap onset relative to the vortex ordering temperature.
In principle these can be completely unrelated, so for our purposes here we will consider the ``worst case scenario" of $T_{\rm BCS} \approx T_{\rm BKT}$~\footnote{Specifically, we will take $T_{\rm BCS} = T_{\rm BCS}^{(0)}$ which in our model corresponds to $T_{\rm BKT} = 0.8117 T_{\rm BCS}$, though our model does not treat the temperature dependence of $T_{\rm BKT}^{(0)}$ and therefore this should be regarded as only a model-dependent example.}.
This corresponds to the case where the two temperature scales are not well-separated, and the quasiparticle coherence peak in the noise can easily be confused with the noise maximum due to vortex motion. 
We also comment that since the superfluid density necessarily drops to zero upon approach $T_{\rm BCS}$ and the physical BKT transition temperature is determined by the relation $\rho_{\rm 2D}^*(T_{\rm BKT}) = 2 T_{\rm BKT}/\pi$, it will always be the case that $T_{\rm BKT} \leq T_{\rm BCS}$. 

We also must in principle compare the characteristic vortex noise scale $\omega_0$ to the relevant quasiparticle frequency scale, which within the Mattis-Bardeen model is referenced with respect to $T_{\rm BCS}$ (or alternatively the zero-temperature gap $\Delta_{\rm BCS}(0)$).
For the sake of simplicity, we will assume that $\omega \ll \omega_0$ such that we can use the low-frequency vortex noise result.
However, it is known that within the Mattis-Bardeen model $\mathrm{Re}(\sigma_{\rm qp}(\omega)) \sim \log \omega$ at low frequencies and therefore we cannot take the $\omega \to 0$ limit in the background. 
For the superconductors we are focused on, we can expect $T_{\rm BCS}$ to correspond to frequencies of order 1THz (1THz$\sim$50K), and therefore frequencies of order $10^{-5}T_{\rm BCS}$ correspond to 10's of MHz, which is within the frequency range we are interested in probing. 
Therefore, we will evaluate the background at frequencies of order $10^{-5}T_{\rm BCS}$.

\begin{figure}
    \centering
    \includegraphics[width=\linewidth]{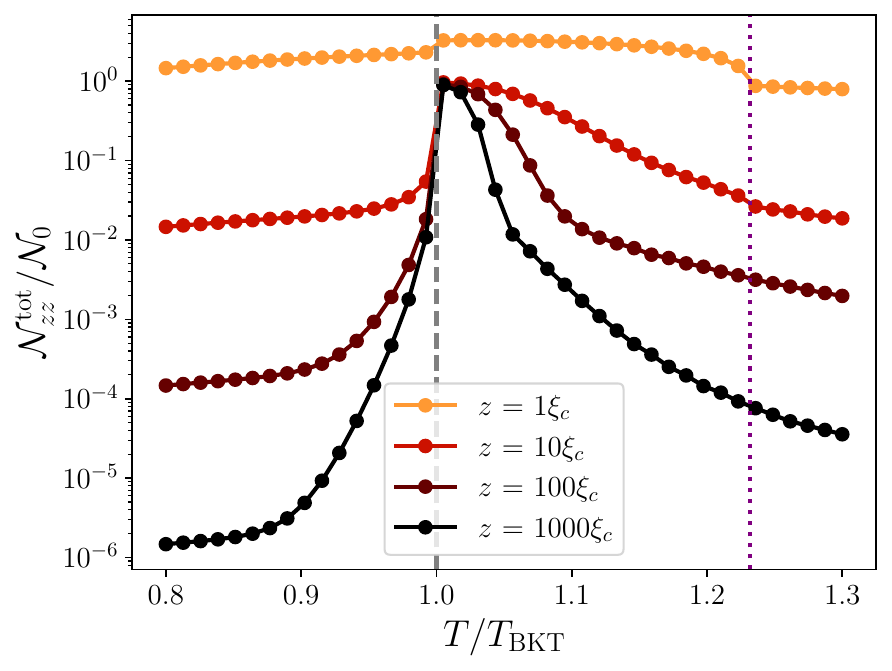}
    \caption{Total magnetic noise at low frequencies including both the vortex and quasiparticle background as a function of temperature for different distances $z$. 
    The dashed gray line is the modeled BKT transition temperature whereas the dotted purple line is the modeled BCS temperature $T_{\rm BCS}$, which we also take to coincide with the unrenormalized BKT temperature $T_{\rm BKT}^{(0)}$. }
    \label{fig:noise-tot}
\end{figure}

In Fig.~\ref{fig:noise-tot} we present the total magnetic noise including both the vortex noise and the quasiparticle background (using the parameters stated above).
We see that even in this case, which was not particularly favorable in terms of $T_{\rm BCS}/T_{\rm BKT}^{(0)}$, there are still clear signatures of the BKT transition in the magnetic noise for all but the closest distances or lowest temperatures.

This can be understood as due to two fortuitous facts.
The first is that the temperature dependence of the quasiparticle background conductivity is slowly varying on the scale of the relevant temperature range where BKT physics is important, so that it is essentially just a constant.
The second is that on the length scales relevant to vortex physics, the quasiparticle response is expected to be local (and therefore momentum independent).
The quasiparticle noise will therefore fall off with distance much faster than the vortex noise which becomes scale invariant at the transition.
We conclude by remarking that in the case of cleaner superconductors, especially such as Van der Waals systems or other low-density clean two-dimensional superconductors, the quasiparticle background noise is not described by the Mattis-Bardeen model, and may potentially be further suppressed.


\section{Conclusion}
\label{sec:conclusion}

We have shown that local spin-qubit magnetometry may be a promising probe for detecting the BKT transition in thin films and two-dimensional superconductors.
In particular, the local magnetic noise gives access to the vortex-antivortex dielectric function $\epsilon_v(\omega,q)$, resolved in both frequency and momentum, thereby enabling one to directly study the transition in terms of the dynamic and scale-dependent vortex interactions.
By analyzing this magnetic noise as a function of temperature, probe frequency, and qubit-sample distance we can identify a number of distinct features of the BKT transition.

Specifically, we observe that the low-frequency magnetic noise is non-monotonic as a function of temperature. The noise maximum, at $T_{\rm BKT}$ exhibits a scaling collapse as a function of the sample-probe distance $z$.
While the magnetic noise obeys a power-law behavior in $z$ below $T_{\rm BKT}$, above the transition temperature we find a more complicated dependence.
This culminates in a plateau structure in the magnetic noise which allows for probing the finite-size crossover effects directly {\it in situ}.
Finally, above the transition it is possible to additionally probe the Drude weight of the free vortices in parallel to the noise from bound-pairs.

The level of detail offered by noise magnetometry can also help address fundamental challenges in identification of BKT physics. 
By probing at short length scales of the order of 10-100$\,$nm, it seems possible to i) disentangle long range disorder effects and ii) use the lateral scanning capability of qubit sensors to study how such disorder affects both the local and global transition properties~\cite{Benfatto.2008,Benfatto.2009}.

We have also explicitly computed the magnetic noise due to Gaussian Aslamazov-Larkin superconducting fluctuations, and shown that there is a clear qualitative \textit{and} quantitative difference between these fluctuations and fluctuations due to vortex unbinding.
While both types of fluctuations manifest similarly in transport quantities and lead to broadening of the superconducting transition, we have shown they contribute differently to the scaling of the magnetic noise with distance.
As a result it may be possible to disentangle these two sources of fluctuations using noise magnetometry.
In addition to the candidate systems listed in Sec.~\ref{sec:expt}, recently the possibility of some two-dimensional materials exhibiting triplet superconductivity has been suggested, in which case the combination of intertwined spin and charge fluctuations~\cite{Curtis.2023,Chung.2022} may exhibit signatures in the local noise magnetometry spectrum.

We also modeled a potential source of background noise due to quasiparticles within the dirty limit using the Mattis-Bardeen model.
Even though this background can potentially be sizeable if $T_{\rm BCS} \sim T_{\rm BKT}$ (i.e. the BKT temperature and pairing temperature are not well separated), we have shown that it should still be possible to clearly observe vortex noise.
This is because the quasiparticle background has a slower-varying temperature dependence and more rapid fall-off with distance (as $1/z^2$) than the vortex noise does.
This points to another key capability\textemdash the ability to probe across different distances\textemdash that spin-qubits such as NV centers offer.

More broadly, while our results are only relevant for superconducting BKT physics, it may also be possible to identify BKT physics in two-dimensional magnetic systems.
In this case, the relation between the magnetic noise and the corresponding vortex dielectric function may be different; however, it still stands that the increased detail and sensitivity may enable further study of BKT physics in a variety of platforms.
Recently, a number of two-dimensional magnets have been shown to exhibit BKT like phase transitions~\cite{Bedoya-Pinto.2021,Klein.2021,Klein.2022,Klein.2022vhg,Augustin.2021,Lu.2020o4,Zhu.2021} which may be amenable to study by noise magnetometry in a similar fashion. 
Interesting features have also recently been seen in the high-symmetry two-dimensional magnetic material CrCl$_3$ using NV noise magnetometry~\cite{Xue.2024}.

Finally, it would be interesting to study the quantum corrections to vortex motion and how they manifest in the local magnetic noise.
In particular, in the presence of strong Coulomb interaction effects the ground-state of a two-dimensional superconducting system should undergo a phase transition between a Mott insulator of Cooper pairs and a superconducting condensate of Cooper pairs. 
This transition can also be described in terms a ``dual" picture based on superconducting vortices, such that a superfluid of vortices corresponds to a Mott insulator of pairs and vice versa, in a conjecture known as the ``particle-vortex duality"~\cite{Fisher.1990,Fazio.2001,Blatter.1994}. 
As we have shown here, it is possible to directly probe the motion of vortices using noise magnetometry; this raises the interesting possibility of directly probing the nature of the particle-vortex duality using spin-qubits.

\begin{acknowledgements}
The authors would like to acknowledge productive discussions with Pavel Dolgirev, Assa Auerbach, Ilya Esterlis, and Johannes Cremer. J.B.C., N.R.P., A.Y., and P.N. are supported by the Quantum Science Center (QSC), a National Quantum Information Science Research Center of  the  U.S.  Department  of  Energy  (DOE). N.R.P. is supported by the Army Research Office through an NDSEG fellowship. N.M. is supported by an appointment to the Intelligence Community Postdoctoral Research Fellowship Program at Harvard University administered by Oak Ridge Institute for Science and Education. A.Y. is partly supported by the Gordon and Betty Moore Foundation through Grant GBMF 9468 and by the ARO Grant W911NF-22-1-0248 P.N. is a Moore Inventor Fellow and gratefully acknowledges support through Grant GBMF8048 from the Gordon and Betty Moore Foundation. 
ED acknowledges support from the ARO grant number W911NF-21-1-0184, the SNSF project 200021$\_$212899, and the Swiss State Secretariat for Education, Research and Innovation (contract number UeM019-1).
B.H. acknowledges support from NSF grant DMR-1231319.
\end{acknowledgements}

\bibliography{references}

\appendix

\section{Magnetostatics}
\label{app:magnetostatic}
In this section, we derive the relation between the magnetic field noise at a distance $z$ from the superconducting sample in terms of the vortex correlation functions.
We assume the superconductivity is truly two-dimensional, i.e. the sample thickness is much smaller than the penetration depth. 

In this case, and in the magnetostatic limit (valid for frequencies $\omega \ll c/z$ which is manifestly realized here), Ampere's law relates the magnetic field $\mathbf{B}$ to the current-density $\mathbf{j}$ via 
\begin{equation}
    \nabla \times \mathbf{B}  = \mu_0\mathbf{j}.
\end{equation}
We have assumed the sample to be modeled as an infinite sheet in the $z= 0$ plane.
In this case, we can characterize the current in terms of a sheet current density via 
\begin{equation}
    \mathbf{j} = \delta(z) \mathbf{j}_\square.
\end{equation}
We perform a Fourier transform on the in-plane coordinates, writing in terms of the two-dimensional in-plane momentum $\mathbf{q}$.
We can use Gauss' law of magnetism and take the curl of this to derive the equation for the $z$-component of the magnetic field which obeys 
\begin{equation}
    \left[ -\partial_z^2 + \mathbf{q}^2 \right]B_z(\mathbf{q},t) = \mu_0 \delta(z) \mathbf{e}_z\cdot i\mathbf{q}\times \mathbf{j}_\square(\mathbf{q},t). 
\end{equation}
This is solved in terms of exponentially decaying solutions as 
\begin{equation}
    B_z(z,\mathbf{q},t) =  B_z(0,\mathbf{q},t) e^{-q|z|}.
\end{equation}
Integrating the singularity across $z=0$ we obtain 
\begin{equation}
\label{eqn:Bz}
    B_z(z,\mathbf{q},t) =  \mu_0 \frac{e^{-q|z|}}{2q} \mathbf{e}_z\cdot i\mathbf{q}\times \mathbf{j}_\square(\mathbf{q},t).
\end{equation}

The local magnetic noise in all components can be reconstructed from knowledge of $B_z$ in conjunction with Maxwell's equations which yields the full vector field as
\begin{equation}
    \mathbf{B}(z,\mathbf{q},t) = \left[ \mathbf{e}_z - \frac{i\mathbf{q}}{q} \right] B_z(z,\mathbf{q},t).
\end{equation}
From this we can compute the magnetic noise tensor at $z > 0$ as 
\begin{multline}
\bm{\mathcal{N}}(z,\omega) =  \int_{\bf q} \int dt e^{i\omega t} \left(\mathbf{e}_z -  i\hat{\mathbf{q}}\right)\otimes\left( \mathbf{e}_z + i\hat{\mathbf{q}}\right) \\
\langle B_z(z,\mathbf{q},t) B_z(z,-\mathbf{q},0)\rangle,
\end{multline}
with $\hat{\mathbf{q}} = \mathbf{q}/q$.
This gives, in terms of the current-current correlation function
\begin{multline}
\label{eqn:noise-tensor}
\bm{\mathcal{N}}(z,\omega) = \frac{\mu_0^2}{4}\int_{\bf q} \left( -i \mathbf{\hat{q}} + \mathbf{e}_z \right)\otimes \left( i \mathbf{\hat{q}} + \mathbf{e}_z \right)\\
e^{-2zq} S^\perp(\mathbf{q},\omega) ,
\end{multline}
where 
\begin{equation}
    S^\perp(\mathbf{q},\omega) = \langle |{j}^\perp_\square(\mathbf{q},\omega)|^2 \rangle 
\end{equation}
is the noise spectral density for the fluctuations of $j_\square^\perp(\mathbf{q},\omega)$, the transverse part of the current fluctuations.
In particular, we will focus on the $zz$ component which is 
\begin{equation}
\label{eqn:noise-zz}
\mathcal{N}_{zz}(z,\omega) = \frac{\mu_0^2}{4}\int_{\bf q} e^{-2zq} S^\perp(\mathbf{q},\omega) .
\end{equation}

In order to proceed further we now specifically consider the case of vortex fluctuations.
In the London-limit, in two-dimensions the supercurrent must be a purely longitudinal response, except in the presence of vortices which introduce topological defects in the phase of the supercurrent. 
In the presence of these vortices, we may express this curl as 
\begin{equation}
    (\nabla \times \mathbf{j}_s)_z = (2e) 2\pi \rho_{\rm 2D}\delta(z) n(\mathbf{r},t),
\end{equation}
where $2e$ is the Cooper pair electric charge, $\rho_{\rm 2D}$ is the {\bf bare} two-dimensional superfluid density, and 
\begin{equation}
    n(\mathbf{r},t) = \sum_j n_j \delta^2(\mathbf{r}-\mathbf{R}_j(t))
\end{equation}
is the two-dimensional vortex density, with $n_j = \pm 1$ and the overall ``neutrality" constraint $\sum_j n_j = 0$ (in analogy with the Coulomb plasma model).
We therefore find 
\begin{equation}
    S^\perp_{\rm BKT}(\mathbf{q},\omega) = (2e)^2 (2\pi \rho_{\rm 2D})^2 \frac{\chi(\mathbf{q},\omega)}{\mathbf{q}^2} ,
\end{equation}
given directly in terms of the vorticity-charge correlation function 
\begin{equation}
    \chi(\mathbf{q},\omega) = \int dt e^{i\omega t} \langle n(\mathbf{q},t) n(-\mathbf{q},0) \rangle   .
\end{equation}
This yields 
\begin{equation}
\mathcal{N}_{zz}(z,\omega) =(2\pi \rho_{\rm 2D} e\mu_0 )^2\int_{\bf q} e^{-2zq} \frac{\chi(\mathbf{q},\omega)}{\mathbf{q}^2}  .
\end{equation}

We now also consider the possibility of a set of $N$ independent 2D superconductor layers equally spaced a distance $a$ apart.
Provided that the coupling between the layers is sufficiently weak (so that it can be ignored over the length scale $N a$), and that the total thickness is still less than the penetration depth, we will see that stacking the layers like this effectively boosts the size of the noise signal, while leaving the other critical physics intact. 
To see this, we simply appeal to the linearity of Maxwell's equations, and superpose the single-layer results to get the total result of 
\begin{equation}
    B_z(z,\mathbf{q},t) =  \sum_{j = 0}^{N-1}\frac{e^{-q|z + j a|}}{2} \mu_0 j_{\square}^\perp(z=-ja,\mathbf{q},t).
\end{equation}
Here we have used coordinates such that the first layer is located at $z = 0$ and then the remaining ones are at $z = -a,-2a, ..., -(N-1)a$, with layer $j$ having a fluctuating vorticity density of $n_j(\mathbf{q},t)$.
Our assumption is that the coupling between the layers is sufficiently weak that these are essentially independently fluctuating quantities. 
We then find the total noise spectrum simply adds in quadrature to give 
\begin{equation}
\mathcal{N}_{zz}(z,\omega) = (\mu_0/2)^2\int_{\bf q} \sum_{j=0}^{N-1} e^{-2|z + j a| q} S^{\perp}(\mathbf{q},\omega) .
\end{equation}
This makes use of $\langle j^\perp_{\square,j}(\mathbf{q},\omega) j^\perp_{\square,k}(-\mathbf{q},-\omega)\rangle \propto \delta_{jk}$.
Now, let us further approximate this sum by 
\begin{equation}
    \sum_{j=0}^{N-1} e^{-2 q j a } = \frac{1 - e^{-2q Na}}{1 - e^{-2qa}}.
\end{equation}
This will, in principle smear out the clarity of the scaling with $z$ due to the fact that a number of depths ranging over $[z,z + Na] $ effectively contribute to the noise simultaneously.
While this is not necessarily detrimental, critical physics will be more straightforwardly observable if the interval is small compared to the overall scale being probed, so that $Na \ll z$.
In this case, we can approximate the summation by 
\begin{equation}
    \frac{1 - e^{-2q Na}}{1 - e^{-2qa}} \sim N + O(a/z)
\end{equation}
using $q \sim 1/z$. 
This then establishes that in this regime of parameters, we have a way of boosting the signal in principle, so that 
\begin{equation}
\label{eqn:noise-zz-layers}
\mathcal{N}_{zz}(z,\omega) = N (2\pi \rho_{\rm 2D} e\mu_0 )^2\int_{\bf q} e^{-2zq} \frac{\chi(\mathbf{q},\omega)}{\mathbf{q}^2}  .
\end{equation}

\section{Extraction of Transition Temperature}
\label{app:BKT-temp}
It is known that it is difficult to extract the exact BKT transition temperature in the thermodynamic limit due to the strong influence of finite size effects. 
In order to isolate the true transition we first numerically integrate the renormalization group equations~\eqref{eqn:rg-eqn}, terminating the flow once the fugacity reaches unity.
This defines the $\xi_+$ length scale by 
\begin{equation}
    y(\xi_+) = 1.
\end{equation}
Near the BKT transition it is known that $\xi_+$ exhibits a divergence as 
\begin{equation}
    \xi_+ =a  \xi_c \exp\left( \frac{b}{\sqrt{T - T_{\rm BKT}}} \right),
\end{equation}
where $a,b>0$ are nonuniversal constants and $T_{\rm BKT}$ can be interpretted as the true transition temperature. 
Numerically, we always truncate our integral at a system-sized infrared cutoff scale of $\ell_{\rm max} \sim 10^{12} \xi_c$, such that in fact the RG flow is truncated at $\min(\xi_+,\ell_{\rm max})$ in practice.
We then perform a linear fit of $1/\log^2(\xi_+/\xi_c)$ versus temperature; the true transition temperature can then be inferred from the $x$-intercept of this fit. 

\begin{figure}
    \centering
    \includegraphics[width=\linewidth]{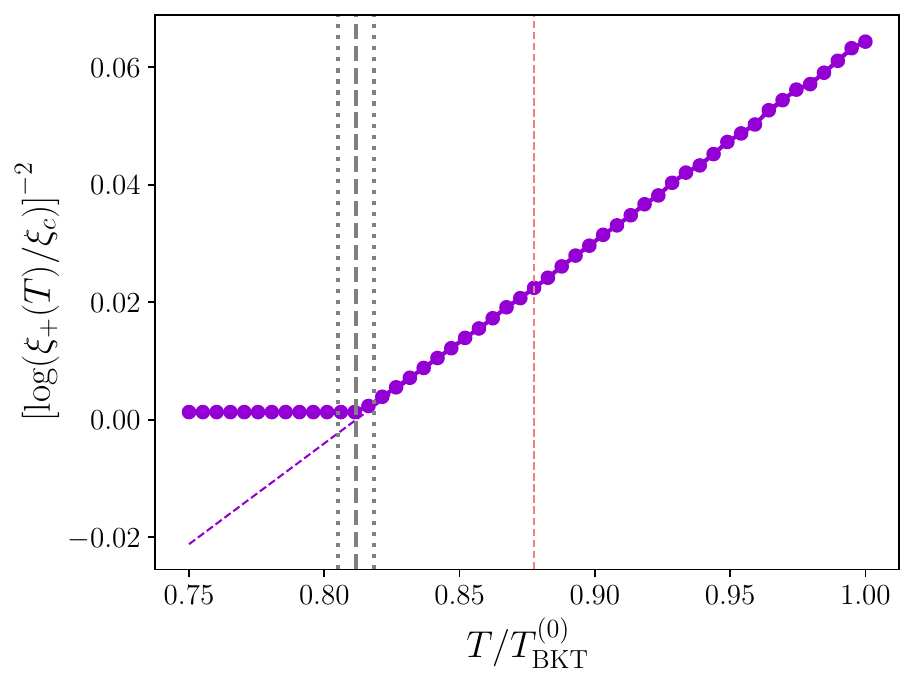}
    \caption{Plot of $1/\log^2(\xi_+/\xi_c)$ versus temperature $T/T_{\rm BKT}^{(0)}$.
    We perform a linear fit based on points with temperature greater than the dotted dashed orange line, which has not yet saturated the finite size cutoff scale. 
    From this we can extract the true transition temperature $T_{\rm BKT}$ (gray dashed line) as the extrapolated $x$-intercept, along with uncertainty estimates (gray dotted lines as 1$\sigma$ interval).
    We find $T_{\rm BKT} = 0.8117 T_{\rm BKT}^{(0)}$, with uncertainty $\Delta T_{\rm BKT} = 0.006T_{\rm BKT}^{(0)}$ with an excellent goodness of fit.}
    \label{fig:tbkt-fit}
\end{figure}

This is shown in Fig.~\ref{fig:tbkt-fit}, from which we extract the true BKT transition temperature in the thermodynamic limit is $T_{\rm BKT} = 0.8117T_{\rm BKT}^{(0)}$ based on our model, along with an uncertainty of $\Delta T_{\rm BKT} = 0.006T_{\rm BKT}^{(0)}$.
It can be seen by eye that the linear fit is excellent in a large temperature regime provided $\xi_+ \lesssim \ell_{\rm max}$; specifically, the goodness of fit for this is over $.9998$, provided we omit the low-temperature points where $\xi_+ \gtrsim \ell_{\rm max}$.

\section{Finite Momentum Scale-Dependent Dielectric}
\label{app:finite-q-eps}
Here we derive the finite-momentum scale-depepdent vortex dielectric function, $\epsilon_v(\mathbf{q})$.
We consider first $T < T_{\rm BKT}$, and compute the bound vorticity density induced by the application of a scalar vorticity potential $\phi$ (here we will not subscript with $v$ to indicate vortex quantities and it will be understood).
In the weakly interacting regime, the vortices are described well by an ensemble distribution function $P(\mathbf{R},\mathbf{r})$ which gives the probability density for a pair to have center-of-mass $\mathbf{R}$ and radial separation $\mathbf{r}$. 
In terms of the free energy this is 
\begin{equation}
    P\propto \exp\left(-\frac{\mathcal{F}(\mathbf{R},\mathbf{r})}{T}\right).
\end{equation}
In addition to the interaction, which is characterized by a running scale-dependent dielectric constant, we also have the local potential, which couples via 
\begin{equation}
    \mathcal{F}_{\rm pot} = - e \rho_{\rm 2D} \left[\phi(\mathbf{R}+\mathbf{r}/2) - \phi(\mathbf{R}-\mathbf{r}/2)\right],
\end{equation}
where we have introduced a fictitious charge $e = 2\pi$ for the vortices.
We compute the induced charge density, which is given by the divergence of the bound polarization density.
This is given, up to linear order in the perturbing potential by 
\begin{equation}
    \mathbf{P}(\mathbf{R}) = \frac{e^2 \rho_{\rm 2D}}{T} \int d^2 r P_0(\mathbf{R},\mathbf{r}) \mathbf{r} \left[ \phi(\mathbf{R}+\mathbf{r}/2) - \phi(\mathbf{R}-\mathbf{r}/2) \right]. 
\end{equation}
We go to momentum space by writing $\phi(\mathbf{r}) = \sum_{\bf q} e^{i\mathbf{q}\cdot\mathbf{r}}\phi(\mathbf{q})$.
This yields 
\begin{equation}
    \mathbf{P}(\mathbf{R}) = \sum_{\bf q} \phi(\mathbf{q}) e^{i\mathbf{q}\cdot\mathbf{R}}\frac{e^2 \rho_{\rm 2D}}{T} \int d^2 r P_0(\mathbf{R},\mathbf{r}) \mathbf{r} 2i\sin(\mathbf{q}\cdot\mathbf{r}/2) . 
\end{equation}
If we assume the equilibrium distribution doesn't depend on center-of-mass coordinate $\mathbf{R}$ we can then express this in momentum space using $\mathbf{P}(\mathbf{R}) = \sum_{\bf q} e^{i\mathbf{q}\cdot\mathbf{R}}\mathbf{P}(\mathbf{q})$ to get   
\begin{equation}
    \mathbf{P}(\mathbf{q}) =  \frac{e^2 \rho_{\rm 2D}}{T} \int d^2 r P_0(\mathbf{r}) \mathbf{r} \left[ 2 i \sin(\mathbf{q}\cdot\mathbf{r}/2) \right]\phi(\mathbf{q}). 
\end{equation}
The induced charge is in turn given by 
\begin{equation}
    \rho_b = -i \mathbf{q}\cdot \mathbf{P}(\mathbf{q}) = 2 \frac{e^2 \rho_{\rm 2D}}{T} \int d^2 r P_0(\mathbf{r}) \mathbf{r}\cdot \mathbf{q}  \sin(\frac{\mathbf{q}\cdot\mathbf{r}}{2}) \phi(\mathbf{q}). 
\end{equation}
We can now obtain from this the full momentum-dependent compressibility as $\kappa = \delta \rho_b /\delta \phi$, giving 
\begin{equation}
    \kappa(\mathbf{q}) = 2 \frac{e^2 \rho_{\rm 2D}}{T} \int d^2 r P_0(\mathbf{r}) \mathbf{r}\cdot \mathbf{q}  \sin(\frac{\mathbf{q}\cdot\mathbf{r}}{2}) . 
\end{equation}
This in turn, through RPA, gives the dielectric constant via 
\begin{equation}
    \epsilon(\mathbf{q}) = 1 + \frac{\kappa(\mathbf{q})}{\mathbf{q}^2}.
\end{equation}

\begin{figure}
    \centering
    \includegraphics[width=\linewidth]{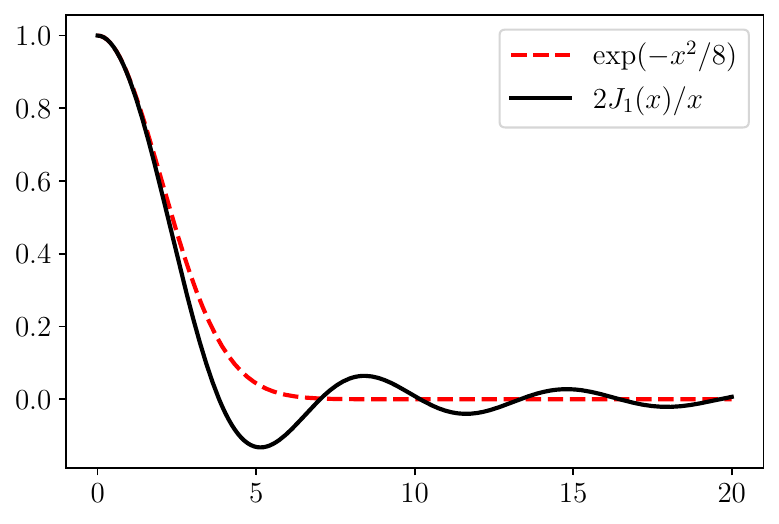}
    \caption{Comparison between true filter function $F(x) = 2 J_1(x)/x$ (black curve) and approximation used here (dashed red curve) of $\tilde{F}(x) = \exp(-x^2/8)$. 
    The approximation is chosen to match the true function up to third order in expansion aroudn $x = 0$. 
    However, the approximation does not the exhibit oscillations which would lead to numerical issues.}
    \label{fig:filter-func}
\end{figure}

In the given expression, we actually have contributions from pairs of all length scales $\mathbf{r}$, and the ultimate result is obtained by summing up all of these contributions. 
We break the total result in to a differential contribution from the shell of pairs between $r \sim \ell$ and $r \sim b\ell$.
This gives the differential contribution to the dielectric of 
\begin{equation}
    d\epsilon(\mathbf{q}) =  2 \frac{e^2 \rho_{\rm 2D}}{T} r dr \int d\theta P_0(\mathbf{r}) \frac{\mathbf{r}\cdot \mathbf{q} }{\mathbf{q}^2} \sin(\frac{\mathbf{q}\cdot\mathbf{r}}{2}).
\end{equation}
Referencing the momentum from angle with $\mathbf{q}$ we find 
\begin{equation}
    d\epsilon(\mathbf{q}) =  4\pi  \frac{e^2 \rho_{\rm 2D}}{T} r^2 dr  \frac{1}{\xi_c^2} P_0(r)  \frac{J_1(q r/2)}{qr} .
\end{equation}
One factor of $1/\xi_c^2$ is simply the areal density available for the center-of-mass distribution.
We note the appearance of the Bessel function $J_1(x)$ due to the angular integral $\int d\theta \cos \theta \sin (x \cos \theta)$.
Let us introduce the filter function 
\begin{equation}
    F(x) = 2 J_1(x)/x.
\end{equation}
This function satisfies $F(0) = 1$ and decays sufficiently rapid for $x \gg 2$.
However, this precise form of the blurring function has issues since it is not positive semi-definite, which can lead to problems when evaluating integrals involving this function numerically. 
We therefore will in practice use a similar function which has the same asymptotic behaviors but is non-negative, such as 
\begin{equation}
    \tilde{F}(x) = \exp(-x^2/a),
\end{equation}
which has the same behaviors at $x = 0,\infty$, while remaining positive and has a smooth maximum at $x = 0$.
We will fix the free constant $a$ by matching the series expansions at $x = 0$ to quadratic order. 
$F(x) = 2 J_1(x)/x \sim 1-  x^2/8$ so we set $a = 8$.
The comparison between these two functions is shown in Fig.~\ref{fig:filter-func}.

In terms of this, we have 
\begin{equation}
    d\epsilon(\mathbf{q}) =  \pi  \frac{e^2 \rho_{\rm 2D}}{T} r^2 dr  \frac{1}{\xi_c^2} P_0(r)  \tilde{F}(q r/2) .
\end{equation}
We can therefore understand this as simply the same scaling equation as the one for the homogeneous scale-dependent dielectric original derived by Kosterlitz and Thouless, weighted by the scale-dependent function $F(q\ell/2)$ which determines how much the momentum $\mathbf{q}$ ends up contributing.
We therefore find the simple result 
\begin{equation}
    \epsilon(\mathbf{q}) = 1 + \int_{\xi_c}^{\infty} d \ell \left( \frac{d\epsilon}{d\ell}\right) \tilde{F}\left(\frac{|\mathbf{q}|\ell}{2}\right).
\end{equation}

Now, we must handle the possibility of free vortices above $T_{\rm BKT}$.
This, however, is essentially already known. 
Above $T_{\rm BKT}$ a new length scale $\xi_+< \infty$ emerges, which signifies the length scale beyond which the fugacity $y \sim 1$ and the perturbative renormalization group fails.
The standard approximation made at this point is to treat all longer length scales as having a finite density of free vortices $n_f = 1/(\pi \xi_+^2)$, which are unbound and respond as free particles to the potential. 
This means that for $T > T_{\rm BKT}$ we should have 
\begin{equation}
    \epsilon(\mathbf{q}) = 1 + \int_{\xi_c}^{\xi_+} d \ell \left( \frac{d\epsilon}{d\ell}\right) \tilde{F}\left(\frac{|\mathbf{q}|\ell}{2}\right) + \frac{4\pi^2 \rho_{\rm 2D}}{\mathbf{q}^2} \frac{n_f}{T} .
\end{equation}
We recognize this as simply the RPA approximation including (i) the contribution from the bound pairs at length scales less than $\xi_+$ and (ii) the free vortices with compressibility $\kappa = n_f/T$ which is expected for a classical Debye fluid.  

In order to capture the dynamic response we will assume that the standard formula involving $\omega$ holds for the bound pairs so that below $T_{\rm BKT}$
\begin{equation}
    \epsilon(\omega,\mathbf{q}) = 1 + \int_{\xi_c}^{\infty} d \ell \left( \frac{d\epsilon}{d\ell}\right) \tilde{F}\left(\frac{|\mathbf{q}|\ell}{2}\right) \frac{14 D \ell^{-2}}{14 D \ell^{-2} - i\omega} ,
\end{equation}
where $D = \mu T$ is the vortex diffusion constant, in terms of the mobility $\mu$.
Above $T_{\rm BKT}$ we will also have to modify the Drude response to take the form 
\begin{multline}
    \epsilon(\omega,\mathbf{q}) = 1 + \int_{\xi_c}^{\xi_+} d \ell \left( \frac{d\epsilon}{d\ell}\right) \tilde{F}\left(\frac{|\mathbf{q}|\ell}{2}\right)\frac{14 D \ell^{-2}}{14 D \ell^{-2}-i\omega} \\
    + \frac{4\pi^2 \rho_{\rm 2D}  n_f \mu }{D\mathbf{q}^2 - i\omega} .
\end{multline}
Here we have used that $D \kappa = \mu n_f$ for the Debye fluid since $\kappa = n_f/T$ and $D = \mu T$. 
We again recall the factor of 14, which while strange, arises from a more precise solution to the bound vortex-pair dynamics in a logarithmic potential~\cite{Ambegaokar.1979}.

\section{Relation to Conductivity}
\label{app:conductivity}
Here we relate the BKT vorticity correlation function to the in-plane transverse conductivity and show how this relates to the standard quantities which figure in to the reflection coefficients, which is an alternative formulation for the noise spectrum. 
The current is given by (in units with $2e = 1$)
\begin{equation}
    \mathbf{J}_s = \rho_{\rm 2D}(\nabla \theta - \mathbf{A}).
\end{equation}
Let us define $\nabla \theta = \mathbf{v}_s$.
The conductivity is found from 
\begin{multline}
    \sigma_{jk}(\omega,\mathbf{q}) = \frac{\rho_{\rm 2D}}{i\omega}\left[ -\delta_{jk} + \frac{\delta \mathbf{v}_s(\omega,\mathbf{q})}{\delta \mathbf{A}(\omega,\mathbf{q})}\right] \\
    = \sigma^{\parallel}(\omega,\mathbf{q}) \frac{q_j q_k}{{q}^2} + \sigma^{\perp}(\omega,\mathbf{q})\left( \delta_{jk}- \frac{q_j q_k}{{q}^2} \right).
\end{multline}

We will focus on the transverse part of the conductivity as this is what contributes to magnetic noise.
For results on the longitudinal conductivity see, e.g. Ref.~\cite{Mikeska.1970}.
To obtain the transverse conductivity, we use the Kubo formula for linear response to obtain
\begin{equation}
    \Re \sigma_{jk}(\omega,\mathbf{q}) = \frac{1}{2T}\Lambda_{jk}(\omega,\mathbf{q})
\end{equation}
with $\Lambda_{jk}(\omega,\mathbf{q})$ the current-current correlation function 
\begin{equation}
    \Lambda_{jk}(\omega,\mathbf{q}) = \langle J^j_s(\omega,\mathbf{q}) J^k_s(-\omega,-\mathbf{q})\rangle. 
\end{equation}
We can separate the current response into longitudinal and transverse parts via 
\begin{multline}
       J^j_s(\omega,\mathbf{q}) = iq_j \rho_{\rm 2D} \theta(\omega,\mathbf{q}) \\
       +  (i\mathbf{q}\times\mathbf{\hat{e}}_z)_j \frac{2\pi \rho_{\rm 2D}}{\mathbf{q}^2} n(\omega,\mathbf{q}).
\end{multline}
We then find for the longitudinal and transverse correlations 
\begin{equation}
    \Lambda^{\parallel}(\omega,\mathbf{q}) = \rho_{\rm 2D}^2 \mathbf{q}^2 \langle|\theta(\omega,\mathbf{q})|^2\rangle,
\end{equation}
and 
\begin{equation}
    \Lambda^{\perp}(\omega,\mathbf{q}) = 4\pi^2 \rho_{\rm 2D}^2 \frac{\langle|n(\omega,\mathbf{q})|^2\rangle}{\mathbf{q}^2}.
\end{equation}
and thus the conductivities 
\begin{subequations}
\begin{align}
    & \Re \sigma^{\parallel}(\omega,\mathbf{q}) = \frac{1}{2T} \rho_{\rm 2D}^2 \mathbf{q}^2 \langle|\theta(\omega,\mathbf{q})|^2\rangle \\
    & \Re \sigma^{\perp}(\omega,\mathbf{q}) = \frac{1}{2T} 4\pi^2 \rho_{\rm 2D}^2 \frac{\langle|n(\omega,\mathbf{q})|^2\rangle}{\mathbf{q}^2} .    
\end{align}
\end{subequations}

To obtain the transverse conductivity we relate the dielectric constant to the vortex-charge response function via 
\begin{equation}
    1- \chi_v(\omega,\mathbf{q}) \frac{4\pi^2 \rho_{\rm 2D}}{\mathbf{q}^2} = \frac{1}{\epsilon_v(\omega,\mathbf{q})}.
\end{equation}
Using the fluctuation dissipation theorem for $\chi_v$ we finally obtain 
\begin{equation}
    \Re \sigma^{\perp}(\omega,\mathbf{q}) = - \frac{\rho_{\rm 2D}}{\omega} \Im \frac{1}{\epsilon_v(\omega,\mathbf{q})}.
\end{equation}
We analytically continue this to 
\begin{equation}
\sigma^{\perp}(\omega,\mathbf{q}) = - \frac{\rho_{\rm 2D}}{i\omega} \frac{1}{\epsilon_v(\omega,\mathbf{q})}.
\end{equation}
This can be understood by taking the zero-temperature result for the transverse response, which is the bare kinetic inductance of $\sigma^\perp = -\rho_{\rm 2D}/i\omega$ and renormalizing the superfluid stiffness to 
\begin{equation}
    \rho_{\rm 2D}^*(\omega,\mathbf{q}) = \frac{\rho_{\rm 2D}}{\epsilon_v(\omega,\mathbf{q})}.
\end{equation}
Below $T_{\rm BKT}$ this can essentially be replaced by the static, long-wavelength dielectric constant so that we simply find the renormalized kinetic inductance of $\rho^*_{\rm 2D} = \rho_{\rm 2D}/\tilde{\epsilon}_v(T)$, with $\epsilon_v(0) = \tilde{\epsilon}_v(T)$ obtained by solving the scaling equations. 
Above $T_{\rm BKT}$, the dielectric constant obtains a singular contribution from the Drude weight of the vortices, so that we can approximate by $\epsilon_v(\omega,\mathbf{q}) \sim \tilde{\epsilon}_v(T_{\rm BKT}^-) + i \gamma/\omega$ at long wavelengths. 
As a result, $\sigma^\perp(\omega) \sim \rho_{\rm 2D}/(i\omega \epsilon_v(\omega)) = \rho_{\rm 2D}/(\gamma - i \omega \rho^*_{\rm 2D}(T_{\rm BKT}^-))$ reflects the onset of a finite resistance in the sample, with conductivity $\rho_{\rm 2D}/\gamma$.

\section{Comparison to Aslamazov-Larkin fluctuations}
\label{app:AL}

Here we present the details of the calculation of the current fluctuations due to the Azlamazov-Larkin superconducting fluctuations. 
We require the correlation function
\begin{equation}
    S_{ab}(\mathbf{q},\omega) = \int dt e^{i\omega t} \langle j^a_\square(\mathbf{q},t) j^b_\square(-\mathbf{q},0) \rangle,
\end{equation}
where the sheet current densities are given by Eq.~\eqref{eqn:AL-current} in the main text.
This can be computed by taking advantage of the fact that the fluctuations are Gaussian. 
We find via Wick's theorem
\begin{multline}
    S_{ab}(\mathbf{q},\omega) = \int dt e^{i\omega t}\int_{\bf p} \left( 2 (2e)\nu_F \xi_c^2\right)^2 p_a p_b \\
    \langle {\psi}_{\mathbf{p}-\mathbf{q}/2}(t)\overline{\psi}_{\mathbf{p}-\mathbf{q}/2}(0)\rangle\langle \overline{\psi}_{\mathbf{p}+\mathbf{q}/2}(t){\psi}_{\mathbf{p}+\mathbf{q}/2}(0) \rangle.
\end{multline}
These correlation functions simply decay in time with the rates 
\begin{equation}
    \Gamma_{\bf k} = \nu_F \Gamma \left[ \xi_c^2 \mathbf{k}^2 + r \right],
\end{equation}
and thus we have 
\begin{equation}
    S_{ab}(\mathbf{q},\omega) =(2e)^2 \int_{\bf p} \frac{2\Gamma_{\bf p}(\mathbf{q})}{\omega^2 +\Gamma_{\bf p}(\mathbf{q})^2 } \left( 2 \nu_F \xi_c^2\right)^2 p_a p_b n_{\mathbf{p}-\mathbf{q}/2}n_{\mathbf{p}+\mathbf{q}/2}
\end{equation}
where we have defined the total rate 
\begin{equation}
    \Gamma_{\bf p}(\mathbf{q}) = \Gamma_{\mathbf{p}+\mathbf{q}/2} + \Gamma_{\mathbf{p}-\mathbf{q}/2}.
\end{equation}
We can compute the equilibrium occupations $n_{\bf k} = \langle |\psi_{\bf k}(0)|^2 \rangle$ using the equipartition result 
\begin{equation}
    n_{\bf k} = \frac{T \Gamma }{\Gamma_{\bf k} }.
\end{equation}
In particular, we find the low-frequency transverse fluctuations depend on the probe momentum $\mathbf{q}$ via 
\begin{equation}
    S^\perp(\mathbf{q}) = (2e)^2\int_{\bf p} \frac{2T^2\Gamma^2\left( 2 \nu_F \xi_c^2\right)^2 p^2 \sin^2\theta }{\Gamma_{\mathbf{p}+\mathbf{q}/2}\Gamma_{\mathbf{p} - \mathbf{q}/2}\left[ \Gamma_{\mathbf{p}+\mathbf{q}/2} + \Gamma_{\mathbf{p}-\mathbf{q}/2}\right] },
\end{equation}
where $\theta$ is the angle of the momentum $\mathbf{p}$ as referenced from the external momentum $\mathbf{q}$. 
Explicitly, we find the result (using $\Gamma = \nu_F^{-1} \tau_{\rm GL}^{-1}$) 
\begin{multline}
    S^\perp_{\rm AL}(\mathbf{q})  = (2e)^2\frac{ 4 T^2 \tau_{\rm GL} }{2\pi} \int_0^{2\pi}\frac{d\theta}{2\pi} \sin^2 \theta \int_0^\infty du u^3 \\
    \frac{1}{\left[ u^2 + x^2 +r \right]\left[ (x^2 + u^2 +r )^2 - (2xu \cos \theta)^2 \right]},
\end{multline}
with $x = q\xi_c/2$ the unitless probe momentum in terms of the microscopic coherence length.

We are confronted with the integral 
\begin{equation}
    I(a,b) = \int_0^{2\pi}\frac{d\theta}{2\pi} \frac{\sin^2\theta}{a^2 - b^2\cos^2\theta },
\end{equation}
where $a = \sqrt{x^2 + u^2 +r }$ and $b = 2x u$. 
We note that $a^2 -b^2 \cos^2 \theta \in [ (x-u)^2 +r , (x+u)^2 +r ]$ and therefore will be nonzero for $r > 0$. 
This can be evaluated using the residue theorem; the result is 
\begin{equation}
    I(a,b) = \frac{1}{b^2}\left[ 1 - \frac{\sqrt{a^2 - b^2}}{a} \right].
\end{equation}
We then arrive at
\begin{multline}
   S^\perp_{\rm AL}(\mathbf{q}) = (2e)^2\frac{ T^2 \tau_{\rm GL} }{2\pi x^2} \\
   \times \int_0^\infty du u \frac{  (x^2 + u^2 + r) - \sqrt{\left[ (u+x)^2 +r \right]\left[ (u-x)^2 +r\right]}}{(u^2 + x^2 +r)^2} .
\end{multline}

The magnetic noise at a distance $z$ is then given by 
\begin{multline}
   \mathcal{N}_{zz}(\mathbf{q}) = \int_{\bf q} e^{-2z q} (e\mu_0)^2\frac{ T^2 \tau_{\rm GL} }{2\pi x^2} \\
   \times \int_0^\infty du u \frac{  (x^2 + u^2 + r) - \sqrt{\left[ (u+x)^2 +r \right]\left[ (u-x)^2 +r\right]}}{(u^2 + x^2 +r)^2} .
\end{multline}
It will be convenient to normalize this by an overall noise scale of 
\begin{equation}
\mathcal{N}_0^{\rm AL} = (e\mu_0/\pi)^2 T_c^2 \tau_{\rm GL}\frac{1}{\xi_c^2}.
\end{equation}
This gives 
\begin{multline}
   \mathcal{N}_{zz}(\mathbf{q})/\mathcal{N}_0^{\rm AL} = \left(1+r\right)^2 \int_{0}^\infty dx e^{-4z/\xi_c x} \frac{ 1 }{x} \\
   \times \int_0^\infty du u \frac{  (x^2 + u^2 + r) - \sqrt{\left[ (u+x)^2 +r \right]\left[ (u-x)^2 +r\right]}}{(u^2 + x^2 +r)^2} ,
\end{multline}
where we have used the definitions that $x = q\xi_c/2$ and $r = T/T_{\rm BCS} - 1$.

\end{document}